\documentclass[a4paper,11pt]{article}
\usepackage{jcappub}
\usepackage{amssymb}
\usepackage{graphicx}
\usepackage{amsmath}
\usepackage{hyperref}

\title{\boldmath Dynamical friction can flip the hierarchical three-body system}

\author[1,2,3]{Li Hu}
\emailAdd{huli21@mails.ucas.ac.cn}
\author[4,2,1]{Rong-Gen Cai}
\emailAdd{cairg@itp.ac.cn}
\author[2,5,*]{Shao-Jiang Wang}
\emailAdd{schwang@itp.ac.cn (Corresponding author)}

\affiliation[1]{School of Fundamental Physics and Mathematical Sciences, Hangzhou Institute for Advanced Study (HIAS), University of Chinese Academy of Sciences (UCAS), Hangzhou 310024, China}
\affiliation[2]{Institute of Theoretical Physics, Chinese Academy of Sciences (CAS), Beijing 100190, China}
\affiliation[3]{University of Chinese Academy of Sciences (UCAS), Beijing 100049, China}
\affiliation[4]{Institute of Fundamental Physics and Quantum Technology, Ningbo University, Ningbo 315211, China}
\affiliation[5]{Asia Pacific Center for Theoretical Physics (APCTP), Pohang 37673, Korea}

\abstract{
In recent years, the long-term effects of non-linear perturbations were found to be important for the evolution of the hierarchical triple system, which, for the central third body of a larger mass, can significantly suppress the occurrences of orbital flip that changes the sign of angular momentum of inner binary. However, as the third-body mass increases significantly, the ambient dark matter spike becomes much more dense, rendering the effect of dynamical friction non-negligible. In this work, we take the dynamical friction into account for the first time in the hierarchical triple system up to the octupole order and find that the suppressed occurrences of orbital flip could be recovered, and as the spike index increases, the number of flips could increase over a period of time; meanwhile, as both the inner and outer semi-major axes increase while keeping their ratio fixed, the number of flips could also increase over the same
number of outer orbital periods, making the detection of orbital flip a potential probe of the dark matter via observations of either electromagnetic waves or gravitational waves.
}

\begin{document}
\maketitle
\flushbottom

\section{Introduction}

The hierarchical triple system is a long-standing topic of interest that consists of three bodies, two of which form an inner binary with a small semi-major axis, while the third body provides gravitational perturbations to the inner binary from a far distance. There are many such astronomical triple systems in the Universe, the most common of which is the Earth-Moon-Sun triple system, in which the Earth-Moon orbit is perturbed by the gravity of the Sun. It can also be that a small binary is perturbed by the gravity of a supermassive black hole at the center of a galaxy.

After writing down the Hamiltonian of a triple system, it is straightforward to find that, when ignoring the kinetic energy term of the center of mass, the leading-order contribution is just two binary problems that are not coupled to each other, which is called the inner and outer binaries separately. In the 1960s, Kozai~\cite{Kozai:1962zz} and Lidov~\cite{Lidov:1962wjn} calculated the gravitational perturbation up to quadrupole order and found that in some cases, the eccentricity and inclination of the inner orbit oscillate violently, and hence there is some kind of exchange between these two parameters. Since this phenomenon was originally discovered by von Zeipel in 1910~\cite{von_Zeipel}, it is also called the von Zeipel-Kozai-Lidov mechanism.

In 2011, Naoz~\cite{Naoz:2011eic} pushed the calculations of gravitational perturbations forward to the octupole order and found that during the evolution process, the projection of the angular momentum of the inner orbit along the direction of the total angular momentum may change a sign, that is, the inner orbit can evolve from the prograde orbiting to retrograde orbiting. This phenomenon is usually called the orbital flip, or simply a flip. After that, various corrections were introduced, including the corrections from hexadecapole~\cite{Will:2017vjc} and higher orders~\cite{10.1093/mnras/stw784}, the post-Newtonian correction and gravitational wave radiations~\cite{Randall:2018qna,Naoz:2012bx}, the spin of the third body as a supermassive black hole~\cite{Fang:2019hir}, to name just a few.

The eccentricity of the orbital flip can be so close to unity, which is the most significant difference from the moderate eccentricity variation allowed by the von Zeipel-Kozai-Lidov mechanism. As for the practical applications, in addition to being used to solve the important problem of hot Jupiter~\cite{Naoz:2011eic}, orbital flip also has important astronomical applications in a series of other hierarchical triple systems, such as stellar systems and compact objects~\cite{Naoz_2016}.

However, in recent years, Luo et al.~\cite{10.1093/mnras/stw475}, Lei et al.~\cite{10.1093/mnras/sty2619}, Will~\cite{Will:2020tri}, Tremaine~\cite{10.1093/mnras/stad1029}, and Li et al.~\cite{Li:2025nmo} pointed out that the usual doubly average computation ignores the nonlinear term of the perturbation, called Brown's Hamiltonian in this paper, which can significantly suppress the occurrences of flip when the mass of the third body is much larger than that of two bodies in the inner orbit. This can be seen that, when expanding the Hamiltonian in terms of a ratio of the semi-major axes between the inner and outer binaries, this nonlinear term is between the octupole and hexadecapole orders, and manifests itself as an azimuthal precession of the eccentricity vector at high eccentricity region~\cite{Klein:2024jxy}. However, when the mass of the third body is much larger than that of the inner binary, the importance of this term will be comparable to or even exceed the octupole order.

It seems that for such a triple system where the binary with small masses is perturbed by a (super) massive black hole, the flip will be difficult to occur. However, as mentioned in Ref.~\cite{Naoz:2011mb}, the occurrence of flip is always accompanied by the eccentricity of the inner orbit being very close to unity. Therefore, if there is a mechanism that can naturally increase the eccentricity of the inner orbit, it may be possible to recover the flip suppressed by Brown's Hamiltonian. 

As shown in Refs.~\cite{Yue:2019ozq,Hu:2023oiu}, the dynamical friction from the dark matter (DM) spike can drive the eccentricity of binary close to unity, hence it is a promising way to save the flip. Note that some may argue that the backreaction from the binary motion may destroy the DM spike, however, for the self-interacting DM, the energy released from orbits can be stored in the isothermal core of the halo, hence the DM spike can still survive the binary motion~\cite{Alonso-Alvarez:2024gdz}. In particular, the effects of DM wind on a binary pulsar around the galactic center were studied in Ref.~\cite{Pani:2015qhr}, producing a characteristic seasonal orbital modulation and a secular change in the orbital period with a magnitude well within various binary-pulsar observations. The rapid spin of pulsars has been considered in~\cite{Li:2021fyp}.

The outline of this paper is as follows. In Sec.~\ref{sec:evolution}, we study the evolution equations up to octupole order without eliminating ascending nodes for the first time as detailed in Sec.~\ref{subsec:Quad&Oct}, where the equations of motion (EOMs) of octupole Hamiltonian are provided in Appendix~\ref{app:octupole}. Meanwhile, we also consider the effects from both Brown's Hamiltonian term and dynamical friction of DM in Sec.~\ref{subsec:BrownHamiltonian} and Sec.~\ref{subsec:DynamicalFriction}, respectively. Finally in Sec.~\ref{sec:results}, we find that the introduction of DM spike can indeed turn the suppressed flip into a successful flip over a large range of dark matter densities. Moreover, as the spike index $\gamma_{\mathrm{sp}}$ increases, the number of flips would increase over a period of time; meanwhile, as both the inner and outer semi-major axes increase while keeping their ratio constant (DM density decreases), the number of flips would also increase within the same number of outer orbital periods due to the greatly extended dynamical friction action time. This can be used to constrain the density of dark matter around (super)massive black holes via astonishing precision of pulsar-timing techniques for binary pulsar systems~\cite{Pani:2015qhr}. Besides, for triple black hole system, the high orbital eccentricity induced by the outer perturber can lead to a more efficient merger rate~\cite{Blaes:2002cs}, accompanied by a stronger and very different spectrum of GW emission~\cite{Yunes:2010sm}. In such cases, the flip can act as a promising probe into DM not only through electromagnetic waves but also gravitational radiations. The last section is devoted to conclusions and discussions.

\section{Evolution of hierarchical triple system}\label{sec:evolution}

\subsection{Quadrupole and Octupole}
\label{subsec:Quad&Oct}

For a given gravitational triple system with a mass hierarchy $m_1\sim m_2\ll m_3$ so that two objects $m_1$ and $m_2$ form a small orbit while the third one $m_3$ is on a much larger orbit, then the evolution of the orbits can be calculated analytically using perturbation method. 

\begin{figure}[htbp]
\centering
\includegraphics[width=0.45\textwidth]{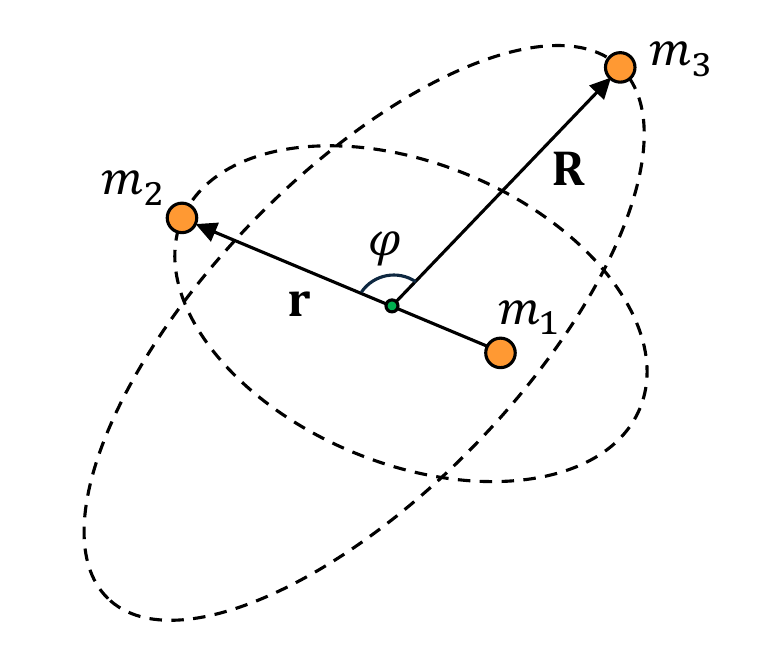}
\caption{A hierarchical triple system consisting of objects with masses $m_1$, $m_2$, and $m_3$ as shown in orange solid circles. The green dot denotes the mass center of $m_1$ and $m_2$. $\bf{r}$ is a distance from $m_1$ to $m_2$. $\bf{R}$ is a distance from the mass center to $m_3$. $\varphi$ is the angle between these two vectors.\label{fig:Triple}}
\end{figure}

Choosing an inertial reference frame, the position vectors of these objects are ${\bf r}_1$, ${\bf r}_2$, and ${\bf r}_3$. 
Let $\bf{r}$ be the relative position vector from $m_1$ to $m_2$, and $\bf{R}$ be the relative position vector from mass center of the inner orbit to $m_3$, as shown in Fig.~\ref{fig:Triple}. The Hamiltonian of the system can be expressed as~\cite{Naoz:2011mb,Randall:2018qna} 
\begin{align}\label{eq:Hamiltonian}
\mathcal{H}=\frac{Gm_1m_2}{2a_1}+\frac{Gm_3(m_1+m_2)}{2a_2}+\sum_{n=2}^{\infty}\frac{Gm_1m_2m_3}{m^n}\left[m_1^{n-1}+(-1)^nm_2^{n-1}\right]\frac{r^n}{R^{n+1}}{\rm P}_n(\cos{\varphi})\,,
\end{align}
where $G$ is the gravitational constant, $a_1$ and $a_2$ denote the semi-major axes (SMAs) of inner orbit and outer orbit, respectively, $\rm{P}_n(x)$ is the Legendre polynomial, and $\varphi$ is the angle between $\bf{r}$ and $\bf{R}$. The motion of the mass center of the three-body system has been neglected in the Hamiltonian since it is trivial. Note also that our Hamiltonian $\mathcal{H}$ is chosen to be positive for the bound orbit, consistent with the convention of Ref.~\cite{Naoz:2011mb}. 



For an elliptical orbit, in addition to the semi-major axis $a_j$, we need the eccentricity $e_j$ and true anomaly $\psi_j$ to further characterize the shape and location of the rotating object, respectively, and the three Euler angles (the longitude of ascending node $h_j$, the inclination $i_j$, and the argument of periastron $g_j$) are also needed to characterize the orientation of the orbit. In the triple system, the subscript $j$ is used to distinguish the inner and outer orbits, and subscript $1$ represents the inner orbit while the subscript $2$ represents the outer orbit.

The determination for the angle $\varphi$ is not as straightforward as it appears, since the usual expressions of $\bf{r}$ and $\bf{R}$ are defined in the orbit coordinate system. Recall that $\hat{{\bf r}}$ in the inner orbit coordinate system can be written as 
\begin{align}
\hat{{\bf r}}_{in}=
\left(
\begin{array}{ccc}
     \cos\psi_1  \\
     \sin\psi_1  \\
     0
\end{array}
\right)\,,
\end{align}
and $\hat{{\bf R}}$ in the outer orbit coordinate system can also be written as 
\begin{align}
\hat{{\bf R}}_{out}=
\left(
\begin{array}{ccc}
     \cos\psi_2  \\
     \sin\psi_2  \\
     0
\end{array}
\right)\,.
\end{align}
Here we need a rotation matrix $\mathbb{R}$ to transform a vector from the inner orbit coordinate system to the outer orbit coordinate system as they may not be on the same plane. After rotating $\hat{{\bf r}}_{in}$ around the three Euler angles $(g_1,i_1,h_1)$, we will obtain the expression of $\hat{{\bf r}}$ in the reference coordinate system,
\begin{align}
\hat{{\bf r}}_{ref}=R_z(h_1)R_x(i_1)R_z(g_1)\hat{{\bf r}}_{in}
\,,
\end{align}
then rotating it again from the reference coordinate system to the outer orbit coordinate system gives rise to
\begin{align}\label{eq:rotation}
\hat{{\bf r}}_{out}&=R_z^{-1}(g_2)R_x^{-1}(i_2)R_z^{-1}(h_2)\hat{{\bf r}}_{ref}
\nonumber\\
&=R_z^{-1}(g_2)R_x^{-1}(i_2)R_z^{-1}(h_2)R_z(h_1)R_x(i_1)R_z(g_1)\hat{{\bf r}}_{in}
\nonumber\\
&=\mathbb{R}\;\hat{{\bf r}}_{in}
\,,
\end{align}
where
\begin{align}
R_z(\theta)=
\left(
\begin{array}{ccc}
     \cos\theta  &\ -\sin\theta &\quad  0  \\
     \sin\theta  &\ \cos\theta &\quad  0  \\
     0  &\ 0 &\quad  1 
\end{array}
\right)\,,\quad
R_x(\theta)=
\left(
\begin{array}{ccc}
     1  &\quad 0 &\  0  \\
     0  &\quad \cos\theta &\ -\sin\theta  \\
     0  &\quad \sin\theta &\  \cos\theta 
\end{array}
\right)\,.
\end{align}
Finally, the angle $\varphi$ can be determined as $\cos\varphi=\hat{{\bf R}}_{out}^{T}\cdot\hat{{\bf r}}_{out}$.

The six parameters defined above for an elliptical orbit are intuitive, but not straightforward to calculate. Fortunately, there exists a set of canonical variables called Delaunay variables that are suitable to describe the periodic motions, and we use them as intermediate variables to make calculations more convenient. These variables include the mean anomalies $l_1$ and $l_2$, the arguments of periastron $g_1$ and $g_2$, the longitudes of ascending nodes $h_1$ and $h_2$, and their conjugate momenta,
\begin{align}\label{eq:conjugate}
L_1=\frac{m_1m_2}{m_1+m_2}\sqrt{G(m_1+m_2)a_1}\,&,\qquad L_2=\frac{m_3(m_1+m_2)}{m_1+m_2+m_3}\sqrt{G(m_1+m_2+m_3)a_2}\,,
\\
G_1=L_1\sqrt{1-e_1^2}\,&,\qquad G_2=L_2\sqrt{1-e_2^2}\,,
\\
H_1=G_1\cos{i_1}\,&,\qquad H_2=G_2\cos{i_2}\,.
\end{align}

With these Delaunay variables, the EOM can be obtained from the following canonical equations,
\begin{align}\label{eq:EOM}
\frac{dL_j}{dt}=\frac{\partial\mathcal{H}}{\partial l_j}\,,\qquad &\frac{dl_j}{dt}=-\frac{\partial\mathcal{H}}{\partial L_j}\,,
\\
\frac{dG_j}{dt}=\frac{\partial\mathcal{H}}{\partial g_j}\,,\qquad &\frac{dg_j}{dt}=-\frac{\partial\mathcal{H}}{\partial G_j}\,,
\\
\frac{dH_j}{dt}=\frac{\partial\mathcal{H}}{\partial h_j}\,,\qquad &\frac{dh_j}{dt}=-\frac{\partial\mathcal{H}}{\partial H_j}\,,
\end{align}
where $j=1,2$ labels the numbering of the inner and outer binaries. Note that the sign of the canonical relations here is also the same as Ref.~\cite{Naoz:2011mb}. 

It is worth noting that, in order to focus on the pure long-term effect, it is usually necessary to do a doubly average to the Hamiltonian as $\left\langle\left\langle\mathcal{H}\right\rangle\right\rangle\equiv\frac{1}{4\pi^2}\int_0^{2\pi}\mathcal{H}\,\mathrm{d}l_1\mathrm{d}l_2$ before bringing the Hamiltonian $\mathcal{H}$ into the canonical equations, that is, to integrate out $l_1$ and $l_2$. After that, Eq.~(\ref{eq:EOM}) is not needed anymore. On the left side, $\frac{dL_j}{dt}=\frac{\partial\left\langle\left\langle\mathcal{H}\right\rangle\right\rangle}{\partial l_j}=0$ since the averaged Hamiltonian is independent of the mean anomalies $l_1$ and $l_2$; On the right side, $l_1$ and $l_2$ as variables have already been integrated out.

We take the Hamiltonian up to the octupole order (that is,  $\left\langle\left\langle\mathcal{H}\right\rangle\right\rangle=\left\langle\left\langle\mathcal{H}_\mathrm{quad}\right\rangle\right\rangle+\left\langle\left\langle\mathcal{H}_\mathrm{oct}\right\rangle\right\rangle$), one may notice that the Newtonian terms have been directly ignored since they are functions of $a_1$, $a_2$; the corresponding canonical variables are $L_1$, $L_2$, which only contribute to the time derivatives of $l_1$, $l_2$ that are, however, already been integrated out.

Note here we do not use the elimination of ascending nodes to simplify subsequent expressions and calculations, since both effects from follow-up $\mathcal{H}_B$ term and DM will break the condition $\dot{h}_1=\dot{h}_2$, which means that $h_1-h_2$ is no longer a constant.

The quadrupole Hamiltonian averaged over $l_1$ and $l_2$ was shown in Eq.(20) of Ref.~\cite{Naoz:2011mb}, which we repeat here for convenience,
\begin{align}\label{eq:Ham2}
\left\langle\left\langle\mathcal{H}_\mathrm{quad}\right\rangle\right\rangle&=\kappa\{\left[1+3\cos(2i_2)\right]\Big(\left[2+3e_1^2\right]\left[1+3\cos(2i_1)\right]+30e_1^2\cos(2g_1)\sin^2(i_1)\Big)
\nonumber\\
&\quad+3\cos(2\Omega)\left[10e_1^2\cos(2g_1)\{3+\cos(2i_1)\}+4(2+3e_1^2)\sin^2(i_1)\right]\sin^2(i_2)
\\
&\quad+12\{2+3e_1^2-5e_1^2\cos(2g_1)\}\cos(\Omega)\sin(2i_1)\sin(2i_2)
\nonumber\\
&\quad+120e_1^2\sin(i_1)\sin(2i_2)\sin(2g_1)\sin(\Omega)-120e_1^2\cos(i_1)\sin^2(i_2)\sin(2g_1)\sin(2\Omega)\}\,,
\nonumber
\end{align}
with $\kappa=\frac{Ga_1^2m_1m_2m_3}{128a_2^3(m_1+m_2)(1-e_2^2)^{3/2}}$, $\Omega=h_1-h_2$. Substituting $\left\langle\left\langle\mathcal{H}_\mathrm{quad}\right\rangle\right\rangle$ into the EOM, the evolution of orbital variables can be obtained

\begin{align}
\frac{{\rm d}e_1}{{\rm d}t}&=\frac{15a_1^2e_1\sqrt{1-e_1^2}Gm_3}{32a_2^3(1-e_2^2)^{3/2}\sqrt{Ga_1(m_1+m_2)}}
\\
&\quad\times\Big\{4\left[\cos(i_1)\sin^2(i_2)\sin(2\Omega)-\sin(i_1)\sin(2i_2)\sin(\Omega)\right]\cos(2g_1)
\nonumber\\
&\quad+\sin(2g_1)\left[\right.(1+3\cos\{2i_2\})\sin^2(i_1)+\cos(2\Omega)(3+\cos\{2i_1\})\sin^2(i_2)
\nonumber\\
&\quad-2\cos(\Omega)\sin(2i_1)\sin(2i_2)\left.\right]\Big\}\,,\nonumber
\end{align}

\begin{align}
\frac{{\rm d}i_1}{{\rm d}t}&=\frac{-3a_1^2Gm_3}{32a_2^3\sqrt{1-e_1^2}(1-e_2^2)^{3/2}\sqrt{Ga_1(m_1+m_2)}}
\\
&\quad\times\Big\{\sin(i_1)\Big[5e_1^2\cos(i_1)\sin(2g_1)\{\left[3+\cos(2\Omega)\right]\cos(2i_2)+2\sin^2(\Omega)\}
\nonumber\\
&\quad-2\left[2+3e_1^2+5e_1^2\cos(2g_1)\right]\sin(2\Omega)\sin^2(i_2)\Big]-2\sin(2i_2)[5e_1^2\cos(\Omega)\cos(2i_1)\sin(2g_1)
\nonumber\\
&\quad+(2+3e_1^2+5e_1^2\cos\{2g_1\})\cos(i_1)\sin(\Omega)]\Big\}\,,
\nonumber\\
\frac{{\rm d}g_1}{{\rm d}t}&=\frac{-3a_1^2Gm_3\csc(i_1)}{64a_2^3\sqrt{1-e_1^2}(1-e_2^2)^{3/2}\sqrt{Ga_1(m_1+m_2)}}
\\
&\quad\times\Big\{\sin(i_1)\Big[(1+3\cos\{2i_2\})\Big(-3-2e_1^2+5(-1+2e_1^2)\cos(2g_1)-10\cos(2i_1)\sin^2(g_1)\Big)
\nonumber\\
&\quad+2\cos(2\Omega)\sin^2(i_2)\Big(-1+6e_1^2+5(-3+2e_1^2)\cos(2g_1)+10\cos(2i_1)\sin^2(g_1)\Big)\Big]
\nonumber\\
&\quad+4\cos(i_1)\cos(\Omega)\sin(2i_2)\Big[-\Big((-1+e_1^2)(-3+5\cos\{2g_1\})\Big)+10\cos(2i_1)\sin^2(g_1)\Big]
\nonumber\\
&\quad+10\sin(2g_1)\Big[\sin(2i_2)\sin(\Omega)\Big(-2+3e_1^2-(-2+e_1^2)\cos(2i_1)\Big)+(2-e_1^2)\sin(2i_1)\sin^2(i_2)\sin(2\Omega)\Big]\Big\}\,,
\nonumber\\
\frac{{\rm d}h_1}{{\rm d}t}&=\frac{3a_1^2Gm_3\csc(i_1)}{64a_2^3\sqrt{1-e_1^2}(1-e_2^2)^{3/2}\sqrt{Ga_1(m_1+m_2)}}
\\
&\quad\times\Big\{\frac{1}{2}\left[-2-3e_1^2+5e_1^2\cos(2g_1)\right]\sin(2i_1)\Big[2-2\cos(2\Omega)+\cos(2\Omega-2i_2)+6\cos(2i_2)
\nonumber\\
&\quad+\cos(2\Omega+2i_2)\Big]+4[2+3e_1^2-5e_1^2\cos(2g_1)]\cos(\Omega)\cos(2i_1)\sin(2i_2)
\nonumber\\
&\quad+20e_1^2\sin(2g_1)\Big[\sin(2\Omega)\sin(i_1)\sin^2(i_2)+\cos(i_1)\sin(\Omega)\sin(2i_2)\Big]\Big\}\nonumber\,,\\
\frac{{\rm d}e_2}{{\rm d}t}&=0\,,
\\
\frac{{\rm d}i_2}{{\rm d}t}&=\frac{-3a_1^2m_1m_2\sqrt{Ga_2(m_1+m_2+m_3)}\csc(i_2)}{64a_2^4(1-e_2^2)^2(m_1+m_2)^2}
\\
&\quad\times\Big\{40e_1^2\cos(2\Omega)\cos(i_1)\sin(2g_1)\sin^2(i_2)
\nonumber\\
&\quad+2\sin(2\Omega)\sin^2(i_2)\Big[5e_1^2\cos(2g_1)(3+\cos\{2i_1\})+2(2+3e_1^2)\sin^2(i_1)\Big]
\nonumber\\
&\quad+2\sin(2i_2)\Big[-10e_1^2\cos(\Omega)\sin(2g_1)\sin(i_1)+\Big(2+3e_1^2-5e_1^2\cos(2g_1)\Big)\sin(\Omega)\sin(2i_1)\Big]\Big\}\,,
\nonumber\\
\frac{{\rm d}g_2}{{\rm d}t}&=\frac{3a_1^2m_1m_2\sqrt{Ga_2(m_1+m_2+m_3)}\csc(i_2)}{128a_2^4(1-e_2^2)^2(m_1+m_2)^2}
\\
&\quad\times\Big\{2[2+3e_1^2-5e_1^2\cos(2g_1)][\cos(i_2)-5\cos(3i_2)]\cos(\Omega)\sin(2i_1)
\nonumber\\
&\quad+[2+3e_1^2-5e_1^2\cos(2g_1)]\cos(2i_1)\sin(i_2)\Big[9-\cos(2\Omega)+5\cos(2i_2)[3+\cos(2\Omega)]\Big]
\nonumber\\
&\quad+20e_1^2[\cos(i_2)-5\cos(3i_2)]\sin(2g_1)\sin(i_1)\sin(\Omega)
\nonumber\\
&\quad+[2+3e_1^2+15e_1^2\cos(2g_1)]\sin(i_2)[3+\cos(2\Omega)+10\cos(2i_2)\sin^2(\Omega
)]
\nonumber\\
&\quad+10e_1^2\cos(i_1)\sin(2g_1)\sin(2\Omega)[-7\sin(i_2)+5\sin(3i_2)]\Big\}\,,
\nonumber
\end{align}

\begin{align}
\frac{{\rm d}h_2}{{\rm d}t}&=\frac{3a_1^2m_1m_2\sqrt{Ga_2(m_1+m_2+m_3)}\csc(i_2)}{64a_2^4(1-e_2^2)^2(m_1+m_2)^2}
\\
&\quad\times\Big\{4\cos(2i_2)\Big[10e_1^2\sin(2g_1)\sin(\Omega)\sin(i_1)+[2+3e_1^2-5e_1^2\cos(2g_1)]\cos(\Omega)\sin(2i_1)\Big]
\nonumber\\
&\quad+\sin(2i_2)\Big[-2-3e_1^2-3(2+3e_1^2)\cos(2i_1)-20e_1^2\cos(i_1)\sin(2g_1)\sin(2\Omega)
\nonumber\\
&\quad+4\cos(2\Omega)\sin^2(i_1)+6e_1^2\cos(2\Omega)\sin^2(i_1)+5e_1^2\cos(2g_1)[\cos(2\Omega)(3+\cos\{2i_1\})-6\sin^2(i_1)]\Big]\Big\}\nonumber\,,
\end{align}
which is similar to Eq.(19) in Ref.~\cite{Fang:2019hir}.

The reader may have noticed that the time derivative of $e_2$ is zero! This is because Hamiltonian $\left\langle\left\langle\mathcal{H}_\mathrm{quad}\right\rangle\right\rangle$ is independent of $g_2$, causing the angular momentum of the outer orbit $G_2$ will not change during the evolution process. Then, considering $G_2=L_2\sqrt{1-e_2^2}$, since $L_2$ is a constant, it is easy to conclude that $e_2$ is a constant.

Next, we move to the octupole order. We define $B=2+5e_1^2-7e_1^2\cos(2g_1)$ and $A=(4+B-4e_1^2)\cos(2i_1)$, and then we give rise to the doubly averaged octupole Hamiltonian $\left\langle\left\langle\mathcal{H}_\mathrm{oct}\right\rangle\right\rangle$ as below without the elimination of ascending nodes for the first time,
\begin{align}\label{eq:Ham3}
\left\langle\left\langle\mathcal{H}_\mathrm{oct}\right\rangle\right\rangle=\alpha\left[C_0+C_1\cos(\Omega)+C_2\cos(2\Omega)+C_3\cos(3\Omega)+C_4\sin(\Omega)+C_5\sin(2\Omega)+C_6\sin(3\Omega)\right]\,,
\end{align}
where
\begin{align}
\alpha&=\frac{15a_1^3e_1e_2Gm_1m_2m_3(m_1-m_2)}{4096a_2^4(1-e_2^2)^{5/2}(m_1+m_2)^2}\,,
\nonumber\\
C_0&=2\sin(g_1)\sin(g_2)\sin(i_1)\left[\sin(i_2)+5\sin(3i_2)\right](-28+5B-56e_1^2-5A)\,,
\nonumber\\
C_1&=-2\cos(g_1)\cos(g_2)\left[16-5B+12e_1^2+5B\cos(2i_1)\right]\left[3+5\cos(2i_2)\right]
\nonumber\\
&\quad+\sin(g_1)\sin(g_2)\cos(i_1)(4+5B-32e_1^2-5A)\left[\cos(i_2)+15\cos(3i_2)\right]\,,
\nonumber\\
C_2&=10\{-8B\cos(g_1)\cos(g_2)\sin(2i_1)\sin(2i_2)
\nonumber\\
&\quad+\sin(g_1)\sin(g_2)\sin(i_1)\left[\sin(i2)-3\sin(3i_2)\right](-4+3B-24e_1^2+A)\}\,,
\nonumber\\
C_3&=-20\sin^2(i_2)\{\cos(g_1)\cos(g_2)\left[16-5B+12e_1^2-3B\cos(2i_1)\right]
\nonumber\\
&\quad-\cos(i_1)\cos(i_2)\sin(g_1)\sin(g_2)(-20+7B-64e_1^2+A)\}\,,
\nonumber\\
C_4&=2\sin(g_1)\cos(g_2)\cos(i_1)(-4-5B+32e_1^2+5A)\left[3+5\cos(2i_2)\right]
\nonumber\\
&\quad-\cos(g_1)\sin(g_2)\left[16-5B+12e_1^2+5B\cos(2i_1)\right]\left[\cos(i_2)+15\cos(3i_2)\right]\,,
\nonumber\\
C_5&=-80\sin(i_1)\sin(i_2)\{-\sin(g_1)\cos(g_2)\cos(i_2)(-4+3B-24e_1^2+A)
\nonumber\\
&\quad+B\cos(g_1)\sin(g_2)\cos(i_1)\left[1+3\cos(2i_2)\right]\}\,,
\nonumber\\
C_6&=20\sin^2(i_2)\{-\sin(g_1)\cos(g_2)\cos(i_1)(-20+7B-64e_1^2+A)
\nonumber\\
&\quad+\cos(g_1)\sin(g_2)\cos(i_2)\left[-16+5B-12e_1^2+3B\cos(2i_1)\right]\}\,.
\nonumber
\end{align}
Substituting $\left\langle\left\langle\mathcal{H}_\mathrm{oct}\right\rangle\right\rangle$ into the EOM, the corresponding evolution equations can be obtained (see Appendix~\ref{app:octupole} for more details).

\subsection{Brown's Hamiltonian}\label{subsec:BrownHamiltonian}

If $m_3$ is comparable to $m_1+m_2$, considering $\mathcal{H}_{quad}$ and $\mathcal{H}_{oct}$ in terms of gravitational perturbations is sufficient. However, for many astronomical systems, $m_3$ can be much larger than $m_1+m_2$. In this case, nonlinear perturbations are significantly enhanced, comparable to, or greater than the effects of an octupole or even a quadrupole. These long-term effects of nonlinear perturbations can be described by adding an additional Hamiltonian, which is called Brown’s Hamiltonian as shown in Ref.~\cite{10.1093/mnras/stad1029}. Note that these effects from nonlinear perturbations are also called second-order quadrupole ($Q^2$) in Ref.~\cite{Will:2020tri}. We take the form of $\mathcal{H}_B$ from Eq.(64) of Ref.~\cite{10.1093/mnras/stad1029},
\begin{align}\label{eq:HB}
\mathcal{H}_B=&\frac{3Gm_1m_2m_3^2a_1^{7/2}(3+2e_2^2)\cos(i_1+i_2)(1-e_1^2)^{1/2}}{64(m_1+m_2)^{3/2}(m_1+m_2+m_3)^{1/2}a_2^{9/2}(1-e_2^2)^3}
\nonumber\\
&\times\left[1+24e_1^2-(1-e_1^2)\cos^2(i_1+i_2)-15e_1^2\sin^2(i_1+i_2)\sin^2g_1\right]\,.
\end{align}
Since $\mathcal{H}_B$ is independent of the mean anomalies $l_1$ and $l_2$, we have $\left\langle\left\langle\mathcal{H}_B\right\rangle\right\rangle\equiv\mathcal{H}_B$. After substituting $\mathcal{H}_B$ into equations of motion, and noting that $H_1=G_1\cos(i_1)$ should be replaced by $H_1=G_1\cos(i_1+i_2)$, the time evolution of inner orbit elements can be calculated as follows,

\begin{align}
\frac{{\rm d}e_1}{{\rm d}t}&=\frac{45a_1^{3}e_1(1-e_1^2)(3+2e_2^2)G^{1/2}m_3^2}{64a_2^{9/2}(1-e_2^2)^3(m_1+m_2)(m_1+m_2+m_3)^{1/2}}\cos(i_1+i_2)\sin^2(i_1+i_2)\sin(2g_1)\,,
\\
\frac{{\rm d}i_1}{{\rm d}t}&=\frac{-45a_1^{3}e_1^2(3+2e_2^2)G^{1/2}m_3^2}{64a_2^{9/2}(1-e_2^2)^3(m_1+m_2)(m_1+m_2+m_3)^{1/2}}\cos^2(i_1+i_2)\sin(i_1+i_2)\sin(2g_1)\,,
\\
\frac{{\rm d}g_1}{{\rm d}t}&=\frac{9a_1^{3}(3+2e_2^2)G^{1/2}m_3^2}{32a_2^{9/2}(1-e_2^2)^3(m_1+m_2)(m_1+m_2+m_3)^{1/2}}
\\
&\quad\times\cos(i_1+i_2)\{8-8e_1^2+5\left[-1+e_1^2+\cos^2(i_1+i_2)\right]\sin^2(g_1)\}\,,
\nonumber\\
\frac{{\rm d}h_1}{{\rm d}t}&=\frac{3a_1^{3}(3+2e_2^2)G^{1/2}m_3^2}{512a_2^{9/2}(1-e_2^2)^3(m_1+m_2)(m_1+m_2+m_3)^{1/2}}
\nonumber\\
&\quad\times\{4-234e_1^2+30e_1^2\cos(2g_1)+45e_1^2\cos\left[2(g_1-i_1-i_2)\right]
\\
&\quad+12\cos\left[2(i_1+i_2)\right]-102e_1^2\cos\left[2(i_1+i_2)\right]+45e_1^2\cos\left[2(g_1+i_1+i_2)\right]\}\,.
\nonumber
\end{align}

\subsection{Dynamical friction}\label{subsec:DynamicalFriction}

Given that the semi-major axis $a_2$ in the specific cases of subsequent studies is always much smaller than the empirical radius $R_{\mathrm{sp}}$ of the spike model, one can consider a spherically symmetric DM spike around the central (super)massive black hole. For an initial DM halo profile $\rho_{\mathrm{halo}}(r)$, the adiabatic growth of the DM profile around the galactic black hole will lead to the formation of a DM
spike~\cite{Gondolo:1999ef,Eda:2014kra}.
\begin{align}\label{spike}
\rho_\mathrm{spike}(r_\mathrm{ISCO}<R<R_\mathrm{sp})=\rho_\mathrm{sp}\left(\frac{R_\mathrm{sp}}{R}\right)^{\gamma_\mathrm{sp}}\,,
\end{align}
The slope $\gamma_\mathrm{sp}$ of the DM spike is steepened into $2.25\leq\gamma_\mathrm{sp}\leq2.5$ for original slope $\gamma_\mathrm{0}$ within a typical range $0\leq\gamma_\mathrm{0}\leq2$, while we treat it as a free parameter within the range $0\leq\gamma_\mathrm{sp}\leq3$ due to the possible disruptive processes to DM spike that may have occurred in the past~\cite{Eda:2014kra}. The innermost stable circular orbit $r_\mathrm{ISCO}$ of the central massive black hole denotes the cut off of the spike. $R$ is the distance between $m_\mathrm{3}$ and the mass center of inner orbit, as shown in Fig.~\ref{fig:Triple}. For an outer orbit that is slightly off-circular (i.e. the eccentricity $e_2$ is not too large), the distance $R$ can be approximated by the semi-major axis of the outer orbit $a_2$. As for empirical radius $R_\mathrm{sp}$ and normalization constant $\rho_\mathrm{sp}$, they are completely determined by masses $m_1$, $m_2$, $m_3$ and redshift $z$, and the detailed calculation of these two parameters can be found in the Appendix~\ref{Spike}

Since the perturbation effect of the outer orbit is not too strong, the decay of the semi-major axis $a_2$ is minor during the evolution process, and the eccentricity of the outer orbit does not change significantly, hence the center of mass of the inner orbit will move approximately on a sphere of a constant radius $R=a_2$, then, it is feasible to set the parameter $\rho_\mathrm{spike}(R)$ to a constant $\rho_\mathrm{spike}(a_2)$ and denoted by $\rho_\mathrm{DM}$. It should be noted that, although one can directly calculate the parameter $\rho_\mathrm{DM}$, such $\rho_\mathrm{DM}$ is subjected to the uncertainty of the associated parameters $R_\mathrm{sp}$, $\rho_\mathrm{sp}$ and $\gamma_\mathrm{sp}$ in practice.

The two objects in the inner orbit experience dynamical friction as described by the Chandrasekhar's formula~\cite{Chandrasekhar:1943ys},
\begin{align}\label{DF}
\mathrm{\bf F}^\mathrm{DF}_{\eta}=-C^\mathrm{DF}m_{\eta}^2\frac{\tilde{\bf v}_{\eta}}{\tilde{v}_{\eta}^{3}}\,,
\end{align}
where $C^\mathrm{DF}=4\pi\mathrm{G}^2\rho_\mathrm{DM}\lambda$, $\tilde{v}_{\eta}\equiv|\tilde{\bf v}_{\eta}|$ and $\Tilde{\bf v}_\eta={\bf v}_\eta+\mathbb{R}^{-1}{\bf V}$. Here the subscript $\eta$ is provided to distinguish each object in the inner orbit, and subscript $1$ denotes the object with mass $m_1$ while subscript $2$ denotes the object with mass $m_2$. $\lambda$ is the Coulomb logarithm, which is defined as $\log(b_{max}/b_{min})$, where $b$ is the impact parameter~\cite{Binney:2008}. It is usually treated as a constant, for example, 3~\cite{Gualandris:2007nm} or 10~\cite{Yue:2019ozq}. In this paper, we take the Coulomb logarithm as $\lambda=10$. $\mathbb{R}$ is the rotation matrix defined in Eq.~(\ref{eq:rotation}). \textbf{V} is the velocity of the mass center of $m_1$ and $m_2$ in the outer orbit coordinate system. One may worry that the back reaction may destroy the DM spike. Nevertheless, for self-interacting DM, the isothermal core of the halo can store the energy released from dynamical friction and hence the DM spike could survive the binary motion~\cite{Alonso-Alvarez:2024gdz}.

In the Chandrasekhar's formula, the orbital velocities of $m_1$ and $m_2$ in the inner orbit coordinate system and the orbital velocity of mass center in the outer orbit coordinate system are given by
\begin{align}
{\bf v}_1&=(-\sqrt{\frac{G\frac{m_2^2}{m_1+m_2}}{a_1(1-e_1^2)}}\sin\psi_1,\sqrt{\frac{G\frac{m_2^2}{m_1+m_2}}{a_1(1-e_1^2)}}(e_1+\cos\psi_1),0)\,,
\\
{\bf v}_2&=(\sqrt{\frac{G\frac{m_1^2}{m_1+m_2}}{a_1(1-e_1^2)}}\sin\psi_1,-\sqrt{\frac{G\frac{m_1^2}{m_1+m_2}}{a_1(1-e_1^2)}}(e_1+\cos\psi_1),0)\,,
\\
{\bf V}&=(-\sqrt{\frac{G(m_1+m_2+m_3)}{a_2(1-e_2^2)}}\sin\psi_2,\sqrt{\frac{G(m_1+m_2+m_3)}{a_2(1-e_2^2)}}(e_2+\cos\psi_2),0)\,,
\end{align}
where $\psi_{1,2}$ are the true anomalies, the relation between the mean anomaly and the true anomaly is
\begin{align}
\mathrm{d}l_j=\frac{(1-e_j^2)^{3/2}}{(1+e_j\cos\psi_j)^2}\mathrm{d}\psi_j\,,\qquad(j=1,2)
\end{align}

Then, the perturbations of the inner and outer orbits from DM can be calculated as
\begin{align}
m_1\ddot{\bf r}_1&={\bf Grav.}+{\bf F}_1^{\rm DF}\,,
\\
m_2\ddot{\bf r}_2&={\bf Grav.}+{\bf F}_2^{\rm DF}\,,
\end{align}
where ${\bf Grav.}$ represents the pure gravitational contribution from the other two bodies. Its effect on the EOM is actually the EOM that we have already obtained from the above quadrupole terms, octopole terms, as well as Brown's Hamiltonian terms~\cite{Will:2020tri}. We abbreviate these contributions as $\bf Grav.$ here because the subsequent calculations focus on the EOM caused by dynamical friction and do not require the specific form of $\bf Grav.$. Here we set ${\bf d}={\bf r}_1-{\bf r}_2$ and ${\bf D}=\frac{m_1{\bf r}_1+m_2{\bf r}_2}{m_1+m_2}-{\bf r}_3$, then we can extract the contribution of DM as ${\bf f}$ and ${\bf F}$ from the following equations,
\begin{align}
\ddot{\bf d}=&{\bf Grav.}+{\bf f}={\bf Grav.}+C^\mathrm{DF}m_{2}\frac{\tilde{\bf v}_{2}}{\tilde{v}_{2}^{3}}-C^\mathrm{DF}m_{1}\frac{\tilde{\bf v}_{1}}{\tilde{v}_{1}^{3}}\,,
\\
\ddot{\bf D}=&{\bf Grav.}+{\bf F}={\bf Grav.}-C^\mathrm{DF}\frac{m_2^2}{m_1+m_2}\mathbb{R}\frac{\Tilde{\bf v}_2}{\tilde{v}_{2}^{3}}-C^\mathrm{DF}\frac{m_1^2}{m_1+m_2}\mathbb{R}\frac{\Tilde{\bf v}_1}{\tilde{v}_{1}^{3}}\,.
\end{align}
The adiabatic-evolution equations for the orbital elements are borrowed from Ref.~\cite{poisson2014gravity},
\begin{align}\label{EOM_DF}
\frac{{\rm d}a_j}{{\rm d}t}&=2\sqrt{\frac{a_j^3(1-e_j^2)^3}{G[m_1+m_2+(j-1)m_3]}}\{e_j\sin(\psi_j)\mathcal{R}_j+\left[1+e_j\cos(\psi_j)\right]\mathcal{S}_j\}\,,
\\
\frac{{\rm d}e_j}{{\rm d}t}&=\sqrt{\frac{a_j(1-e_j^2)}{G[m_1+m_2+(j-1)m_3]}}\{\sin(\psi_j)\mathcal{R}_j+\frac{2\cos(\psi_j)+e_j\left[1+\cos^2(\psi_j)\right]}{1+e_j\cos(\psi_j)}\mathcal{S}_j\}\,,
\\
\frac{{\rm d}i_j}{{\rm d}t}&=\sqrt{\frac{a_j(1-e_j^2)}{G[m_1+m_2+(j-1)m_3]}}\frac{\cos(g_j+\psi_j)}{1+e_j\cos(\psi_j)}\mathcal{W}_j\,,
\\
\frac{{\rm d}g_j}{{\rm d}t}&=\frac{1}{e_j}\sqrt{\frac{a_j(1-e_j^2)}{G[m_1+m_2+(j-1)m_3]}}\{-\cos(\psi_j)\mathcal{R}_j
\nonumber\\
&\quad+\frac{2+e_j\cos(\psi_j)}{1+e_j\cos(\psi_j)}\sin(\psi_j)\mathcal{S}_j-e_j\cot(i_j)\frac{\sin(g_j+\psi_j)}{1+e_j\cos(\psi_j)}\mathcal{W}_j\}\,,
\\
\frac{{\rm d}h_j}{{\rm d}t}&=\csc(i_j)\sqrt{\frac{a_j(1-e_j^2)}{G[m_1+m_2+(j-1)m_3]}}\frac{\sin(g_j+\psi_j)}{1+e_j\cos(\psi_j)}\mathcal{W}_j\,,
\end{align}
where
\begin{align}
\mathcal{R}_1={\bf f}\cdot{\bf n}\,,\qquad\mathcal{S}_1&={\bf f}\cdot{\bf k}\,,\qquad\mathcal{W}_1={\bf f}\cdot{\bf e}_z\,,
\\
\mathcal{R}_2={\bf F}\cdot{\bf N}\,,\qquad\mathcal{S}_2&={\bf F}\cdot{\bf K}\,,\qquad\mathcal{W}_2={\bf F}\cdot{\bf E}_z\,,
\end{align}
and
\begin{align}
{\bf n}&=(\cos(\psi_1),\sin(\psi_1),0)\,,\qquad{\bf k}=(-\sin(\psi_1),\cos(\psi_1),0)\,,\qquad{\bf e}_z=(0,0,1)\,,
\\
{\bf N}&=(\cos(\psi_2),\sin(\psi_2),0)\,,\qquad{\bf K}=(-\sin(\psi_2),\cos(\psi_2),0)\,,\qquad{\bf E}_z=(0,0,1)\,.
\end{align}

Finally, the doubly average should be done for these evolution equations to focus on the long-term effects. Note that since the double integration for the EOM here is too complicated to be solved analytically. In the actual calculation, we use numerical integration instead.

Before showing the results, we want to demonstrate how to combine each part of the equation of motion to obtain a FULL EOM. Here, we take the evolution of the inner orbit eccentricity $e_1$ as an example. It consists of four parts,
\begin{align}
\Big[\frac{{\rm d}e_1}{{\rm d}t}\Big]_\mathrm{FULL}\equiv\Big[\frac{{\rm d}e_1}{{\rm d}t}\Big]_\mathrm{quad}+\Big[\frac{{\rm d}e_1}{{\rm d}t}\Big]_\mathrm{oct}+\Big[\frac{{\rm d}e_1}{{\rm d}t}\Big]_\mathrm{Brown}+\left\langle\left\langle\frac{{\rm d}e_1}{{\rm d}t}\right\rangle\right\rangle_\mathrm{DF}\,,
\end{align}
where $\Big[\cdot\cdot\cdot\Big]$ has no practical meaning, but $\left\langle\left\langle\cdot\cdot\cdot\right\rangle\right\rangle$ denotes doubly average.

\section{Results and analysis}\label{sec:results}

\begin{table}[htbp]
\caption{Parameters and initial conditions selected for illustration. The unit of mass is $M_{\odot}$ and the unit of length is au, hence the unit of DM density is $M_{\odot}/{\rm au}^3$. The distance $a_2$ in both cases ensures that the inner binary is located within the dark matter spike. The Coulomb logarithm $\lambda$ is fixed as 10 in this paper. And the selection of some parameters refers to Tab.~1 of this paper~\cite{Will:2020tri}.}
\label{tab:i}
\centering
\renewcommand{\arraystretch}{1.5}
{\scriptsize
\begin{tabular}{cccccccccccccccc}
\hline
Case & $\gamma_{sp}$ & $\rho_\mathrm{DM}$ & $a_1$ & $a_2$ & $e_1$ & $e_2$ & $m_1$ & $m_2$ & $m_3$ & $g_1$ & $g_2$ & $i_1$ & $i_2$ & $h_1$ & $h_2$\\
\hline
A & 7/3 & $1.35\times10^{-4}$ & 0.2 & 10 & 0.01 & 0.6 & 99 & 1 & 15630 & 30 & 0 & 84 & 0 & 180 & 0 \\
B & 5/3 & $4.26\times10^{-7}$ & 0.1 & 5 & 0.01 & 0.6 & 90 & 10 & 22520 & 0 & 0 & 85 & 0 & 180 & 0 \\
C & 5/3 & $9.19\times10^{-9}$ & 1 & 50 & 0.01 & 0.6 & 90 & 10 & 22520 & 0 & 0 & 85 & 0 & 180 & 0 \\
D & 5/3 & $1.98\times10^{-10}$ & 10 & 500 & 0.01 & 0.6 & 90 & 10 & 22520 & 0 & 0 & 85 & 0 & 180 & 0 \\
E & 5/3 & $4.26\times10^{-12}$ & 100 & 5000 & 0.01 & 0.6 & 90 & 10 & 22520 & 0 & 0 & 85 & 0 & 180 & 0 \\
F & 5/3 & $1.21\times10^{-8}$ & 1 & 60 & 0.01 & 0.45 & 98 & 2 & 100810 & 0 & 0 & 87 & 0 & 180 & 0 \\
G & 2 & $3.02\times10^{-7}$ & 1 & 60 & 0.01 & 0.45 & 98 & 2 & 100810 & 0 & 0 & 87 & 0 & 180 & 0 \\
H & 7/3 & $7.52\times10^{-6}$ & 1 & 60 & 0.01 & 0.45 & 98 & 2 & 100810 & 0 & 0 & 87 & 0 & 180 & 0 \\
I & 2.4 & $1.43\times10^{-5}$ & 1 & 60 & 0.01 & 0.45 & 98 & 2 & 100810 & 0 & 0 & 87 & 0 & 180 & 0 \\
J & 7/3 & $7.53\times10^{-6}$ & 1 & 60 & 0.01 & 0.45 & 98 & 2 & 101000 & 0 & 0 & 87 & 0 & 180 & 0 \\
\hline
\end{tabular}
}
\end{table}

\begin{figure}[htbp]
\centering
\includegraphics[width=\textwidth]{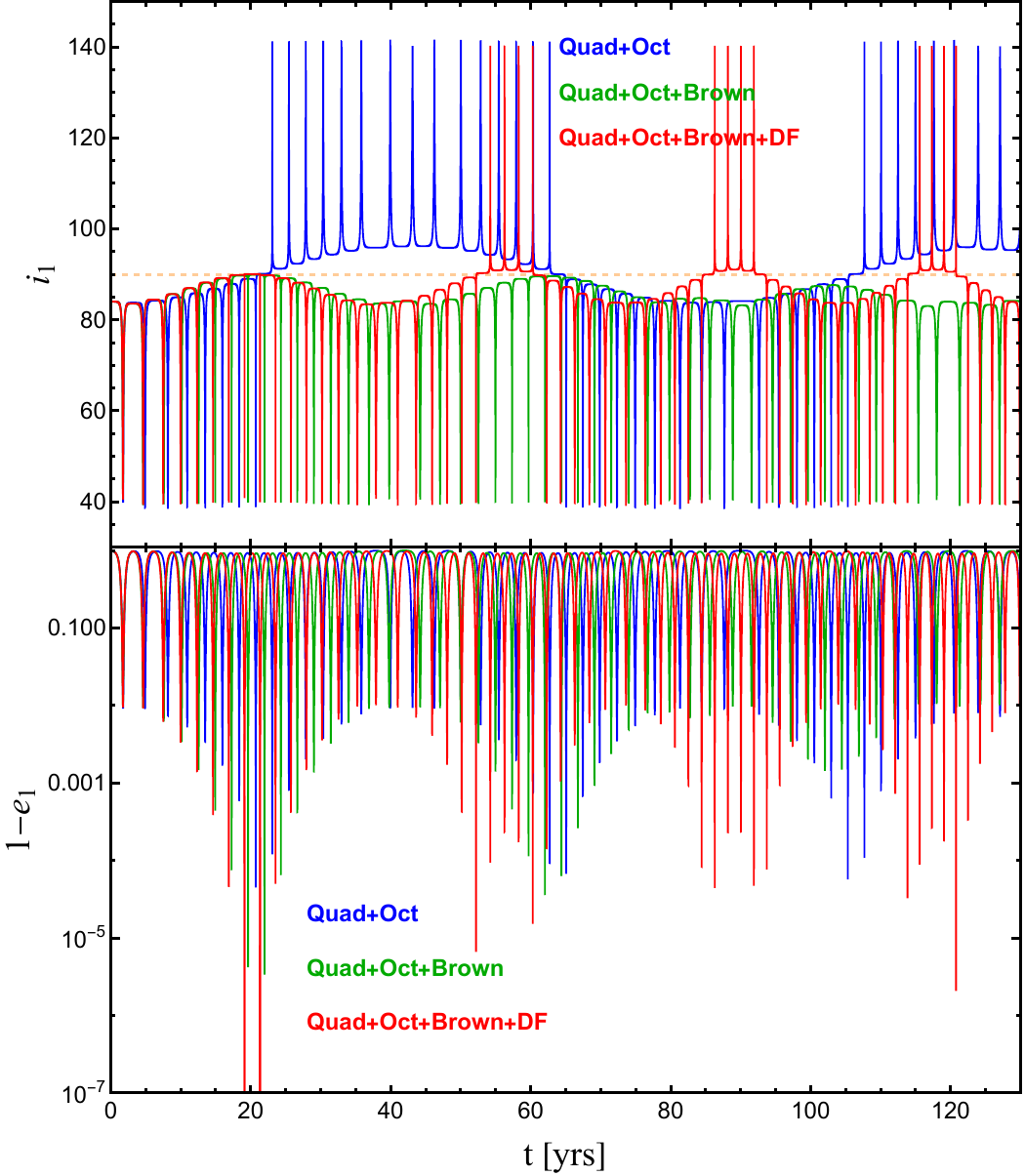}
\caption{The evolution of orbital flips (top) and eccentricity excursions (bottom) in a hierarchical triple system of masses $99M_{\odot}$, $1M_{\odot}$, and $15630M_{\odot}$ with the inclusions of quadrupole and octupole orders (blue), the additional inclusion of Brown's Hamiltonian term (green), and the additional inclusions of both Brown's Hamiltonian term and dynamical friction from a DM spike (red).}
\label{fig:15630}
\end{figure}

For a given hierarchical triple system, if we consider perturbations up to the octupole order, the inner orbit can sometimes flip in its orientation. However, as pointed out by Will~\cite{Will:2020tri}, Luo et al.~\cite{10.1093/mnras/stw475}, and others, the effects of Brown's Hamiltonian can suppress the appearance of flip, which is manifested by a decrease in the duration of flip-up state or even the complete disappearance of flip. This phenomenon can be shown in the top panel of Fig.~\ref{fig:15630} between the blue and green cases, where the corresponding parameters are displayed from an illustrative case A in Tab.~\ref{tab:i}. It is straightforward to see that the flip is suppressed by Brown's Hamiltonian completely, where the curve turns from a flip-up state (blue) to no flip (green) after both 20 years and 105 years.

However, as also seen from the red case in the top panel of Fig.~\ref{fig:15630}, if the effect from dynamical friction is included, the flip could be occasionally recovered, which suggests that the existence of DM can contribute to the occurrences of flip. Conversely, the detection of such a flip can be an important probe into the presence of DM, especially for those astronomical systems that should not flip under the action of gravity but have been observed to flip. Therefore, in such cases, the flip becomes a ''magnifying glass'' for DM detection. It is worth noting that the inner binary is located within the dark matter spike in all cases in Tab.~\ref{tab:i}. This amplification effect is important not only for the detections using electromagnetic waves but also for the detections with gravitational waves~\cite{Naoz:2012bx} since the occurrences of flip is often accompanied by extremely large eccentricities of the inner orbits as shown in the bottom panel of Fig.~\ref{fig:15630}, which necessarily leads to violent gravitational radiations. A similar observational perspective was also proposed in Ref.~\cite{Hu:2023oiu} for eccentric inspirals of supermassive black hole binaries with a DM spike.


Meanwhile, the correspondence between flip and eccentricity does not seem that good in the bottom panel of Fig.~\ref{fig:15630}. Note that the eccentricity of the red curve around $t=20$ yrs in the bottom panel (slightly overlapping with the green curve) is comparable to or even greater than that around $t=60$, $90$, and $120$ yrs, however, there is no flip at around $t=20$ yrs, while there are flips around $t=60$, $90$, and $120$ yrs. This simply indicates that the interchange between the inclination and eccentricity is slightly disrupted.

\begin{figure}[htbp]
\centering
\includegraphics[width=\textwidth]{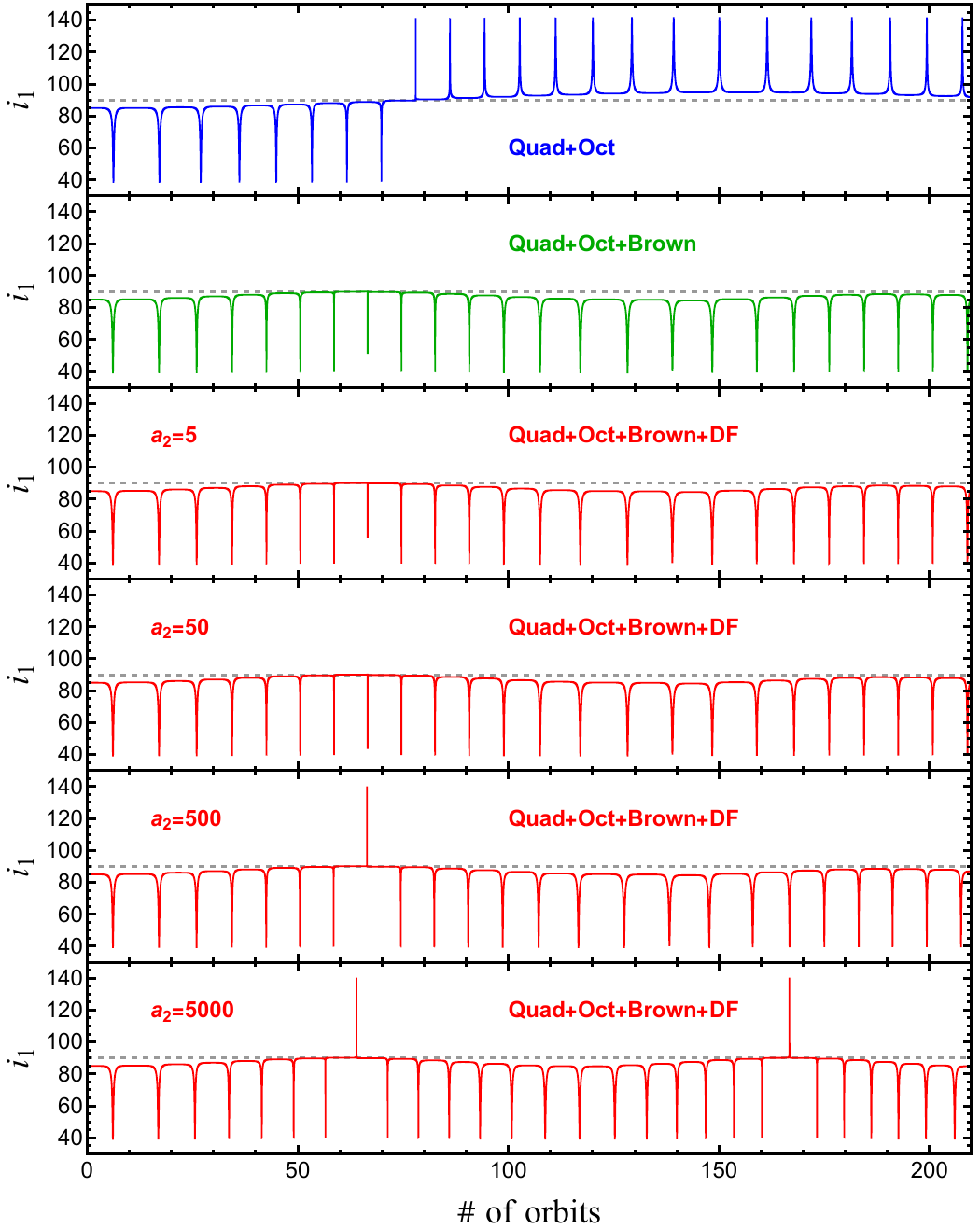}
\caption{The evolution of orbital flips in a hierarchical triple system under parameter sets in cases B, C, D, and E with the inclusions of quadrupole and octupole orders (blue), the additional inclusion of Brown's Hamiltonian term (green), and the additional inclusions of both Brown's Hamiltonian term and dynamical friction from a DM spike (red). Note that the horizontal axis is no longer yrs but is converted to the number of outer orbital periods at the respective distances.}
\label{fig:22520}
\end{figure}

\begin{figure}[htbp]
\centering
\includegraphics[width=\textwidth]{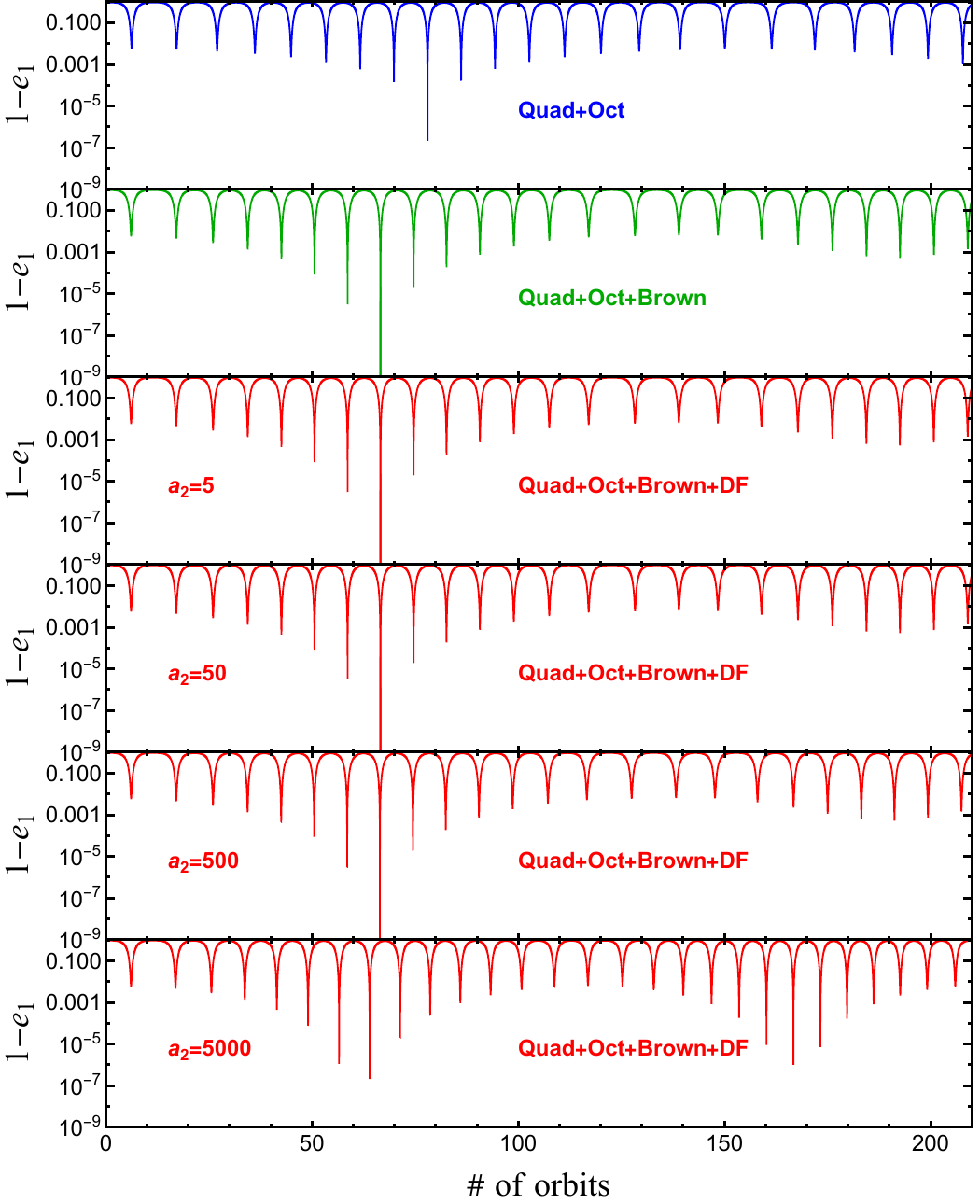}
\caption{The evolution of orbital flips in a hierarchical triple system under parameter sets in cases B, C, D, and E with the inclusions of quadrupole and octupole orders (blue), the additional inclusion of Brown's Hamiltonian term (green), and the additional inclusions of both Brown's Hamiltonian term and dynamical friction from a DM spike (red). Note that the horizontal axis is no longer yrs but is converted to the number of outer orbital periods at the respective distances.}
\label{fig:22520v2}
\end{figure}

Now we address the parameter choice in case A. One of the most important issues is whether the occurrence of orbital flips caused by dynamical friction depends exclusively on a high dark matter density. The answer turns out to be negative. Note that for the blue and green curves, the horizontal time axis can be re-scaled to accommodate astronomical observations by changing the initial values of $a_1$ and $a_2$ with an equal proportionality. A fixing $a_1/a_2$ also fixes the period ratio $P_\mathrm{in}/P_\mathrm{out}$ of the inner and outer orbits. Then, keeping $a_1/a_2=1/50$ fixed and varying $a_1$, $a_2$ at the same time, the DM density $\rho_{\mathrm{DM}}\equiv\rho_{\mathrm{spike}}(a_2)$ would decrease as the semi-major axis $a_2$ increases. The corresponding parameters are displayed from illustrative cases B-E in Tab.~\ref{tab:i}, and the results of case B-E are shown in Fig.~\ref{fig:22520} and Fig.~\ref{fig:22520v2}. As the semi-major axes increase, the evolution curve of $i_1$ in the absence of dark matter remains unchanged with the number of outer orbital periods as the horizontal axis. However, in the presence of dark matter spike, flips could occur and the number of flips could increase as the semi-major axes $a_1$ and $a_2$ increase. When $a_2=5\,\mathrm{au}$ and $50\,\mathrm{au}$, there is no flip within 200 outer orbits; When $a_2=500\,\mathrm{au}$, there is a flip after around 65 outer orbits; When $a_2=5000\,\mathrm{au}$, there are two flips after 65 and 170 outer orbits. Even in the case E shown at the bottom of Fig.~\ref{fig:22520}, where DM density has reached the order as low as $10^{-12}$$M_{\odot}/{\rm au}^3$, flips can still occur. This phenomenon can be explained by the fact that as the semi-major axis $a_2$ increases, although the DM density decreases, the outer orbital period becomes longer, and within the same number of outer orbital periods, the effects from dynamical friction can accumulate for a much longer time, thereby increasing the number of flips within the same number of outer orbital periods. This is also an important manifestation that the angular momentum of the inner and outer orbits is no longer conserved after considering dynamical friction.
 
The same conclusion as in Fig.~\ref{fig:15630} can be obtained for Fig.~\ref{fig:22520v2}: the occurrence of a flip must correspond to an extremely large eccentricity $e_1$, but an extremely large eccentricity $e_1$ does not necessarily correspond to the occurrence of a flip.

\begin{figure}[htbp]
\centering
\includegraphics[width=\textwidth]{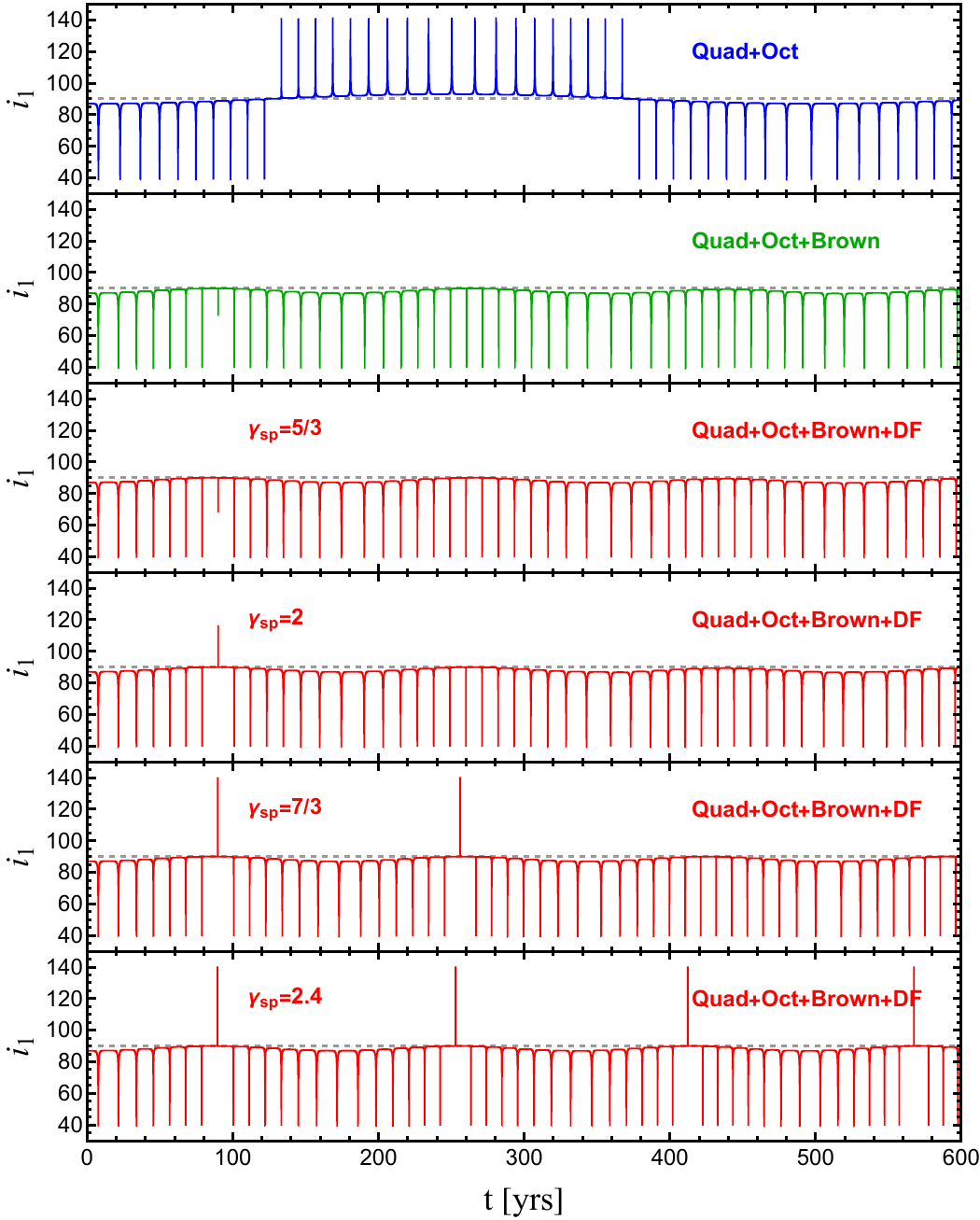}
\caption{The evolution of orbital flips in a hierarchical triple system under parameter sets in cases F, G, H, and I with the inclusions of quadrupole and octupole orders (blue), the additional inclusion of Brown's Hamiltonian term (green), and the additional inclusions of both Brown's Hamiltonian term and dynamical friction from a DM spike (red). Note that the parameters of these six curves differ only in the spike index $\gamma_{\mathrm{sp}}$ and dark matter density $\rho_{\mathrm{DM}}$, the top two (blue and green) are independent of dark matter-related parameters.}
\label{fig:Vary_Density}
\end{figure}

Furthermore, in order to analyze the influence of dynamical friction in detail, we studied the performance of flip under different values of spike index $\gamma_{\mathrm{sp}}$ (hence also different dark matter density), which is displayed in Fig.~\ref{fig:Vary_Density} and the corresponding parameters are shown in Tab.~\ref{tab:i} for the cases F, G, H, and I. It can be seen that there is a long-time flip-up state (blue) when only considering the quadrupole and octupole orders, and again, the inclusion of Brown's Hamiltonian term (green) strongly suppresses the would-be flip-up state. When the spike index of dark matter is relatively small ($\gamma_{\mathrm{sp}}=5/3$), the flips cannot be recovered successfully, which is almost similar to the expected case where the dynamical friction vanishes when there is no dark matter density at all in the previous case (green).

What is fascinating is that as the spike index $\gamma_{\mathrm{sp}}$ increases, the number of flips increases significantly so that the total duration of the flip-up state in the same period of time also increases. When $\gamma_{\mathrm{sp}}=2$, there is only one flip; when $\gamma_{\mathrm{sp}}=7/3$, there are two flips; and when $\gamma_{\mathrm{sp}}=2.4$, there are four flips, which means that as the spike index $\gamma_{\mathrm{sp}}$ of dark matter increases, the DM density $\rho_{\mathrm{DM}}$ will also increase. According to formula Eq.~(\ref{DF}), dynamical friction is significantly enhanced and the eccentricity of the inner orbit is more likely to tend to unity, resulting in an increase in the number of flips over a period of time.

Nevertheless, a full investigation for the parameter space with rescued flip by DM is computationally expensive and hence practically difficult for two reasons: (i) the parameter space is considerably large as it concerns not only the masses but also the relative positions for a triple system with not only eccentricity but also DM environment; (ii) the appearance of a flipped state depends on not only the time resolution (as some of the flipped states can only last for a short period of time) but also the running time (as some of the flipped state can only appear after a long period of time), both of which and their combined would require a huge computational power to achieve a full investigation of parameter space. We will leave this task for a future study, but only present below some illustrative examples of principles to see the feasibility of our findings.

\begin{figure}[htbp]
\centering
\includegraphics[width=\textwidth]{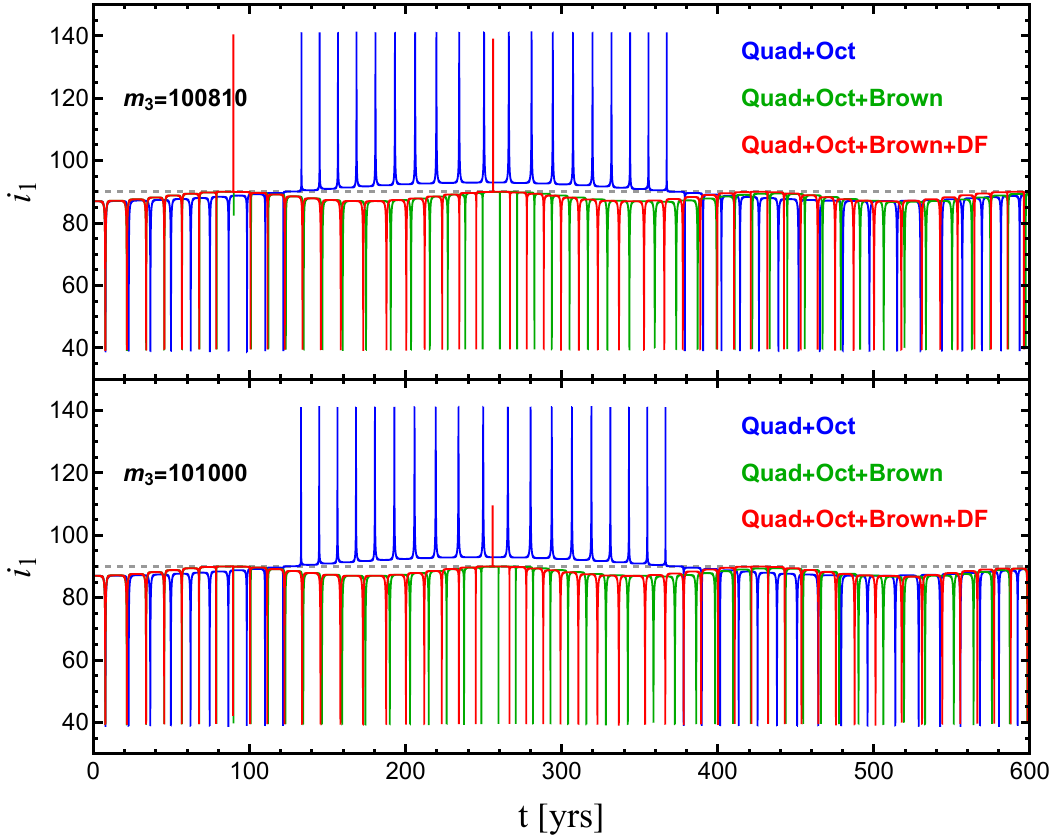}
\caption{The evolution of orbital flips in a hierarchical triple system under parameter sets in cases I (top) and J (bottom) with the inclusions of quadrupole and octupole orders (blue), the additional inclusion of Brown's Hamiltonian term (green), and the additional inclusions of both Brown's Hamiltonian term and dynamical friction from a DM spike (red).}
\label{fig:Vary_m3}
\end{figure}

Fig.~\ref{fig:Vary_m3} is provided to see moderate fine-tuning of $m_3$ to achieve an orbital flip. The critical value of $m_3$ at which the flip is completely suppressed is around $100800\,M_{\odot}$, which is displayed in the top panel of Fig.~\ref{fig:Vary_m3}. After increasing the value of $m_3$ by about $200\,M_{\odot}$, the bottom panel of the figure is achieved. For the situation without dynamical friction (green curve), the flip is suppressed more obviously, hence there is no flip. However, for the case with dynamical friction (red curve), there still comes with a flip although the number of flips is reduced from two to one, suggesting a moderate fine-tuning of $m_3$.

Before concluding, it is necessary to clarify the selection of parameters shown in Tab.~\ref{tab:i}. First, the parameter $\lambda$ is the Coulomb logarithm, which is usually regarded as a constant and is taken to be 10 in our paper~\cite{Binney:2008}. Then, the spike index $\gamma_{\mathrm{sp}}$ is selected between $5/3$ and $2.4$ to investigate the flip response. After that, the DM density $\rho_{\mathrm{DM}}\equiv\rho_{\mathrm{spike}}(a_2)$ is fixed by the DM spike model. In addition, the choice of the semi-major axes $a_1$, $a_2$, eccentricities $e_1$, $e_2$ and the masses $m_1$, $m_2$, $m_3$ is quite arbitrary, but should at least satisfy the requirement that the quadrupole moment perturbation term is no greater than the two leading order terms in Eq.~(\ref{eq:Hamiltonian}), i.e., the weak coupling conditions between the inner and outer orbits are satisfied~\cite{Randall:2018qna},
\begin{align}\label{eq:coupling}
\frac{m_3}{m_1+m_2}\Big(\frac{r}{R}\Big)^3\ll1\,,\qquad\qquad\frac{m_1m_2}{(m_1+m_2)^2}\Big(\frac{r}{R}\Big)^2\ll1\,.
\end{align}
Besides this, to ensure the occurrence of flip at the octupole order, both $m_1-m_2\neq0$ and $e_2\neq0$ must be satisfied~\cite{Naoz:2011mb}. Furthermore, in terms of the choice of $m_3$, except for case J, we choose slightly special values for the others to highlight the contribution of dynamical friction, which is close to the critical value of flip occurrence (not mandatory, as expressed in Fig.~\ref{fig:Vary_m3}). Finally, the choice of $i_1$ is close to $90^\circ$ to make the occurrence of flip easier.

Back to the previous question: one may worry that the DM density is too low in the real universe such that dynamical friction may not play a significant role in the orbital flip. The key point is that the density of dark matter around (super)massive black holes in the spike model could be much higher than the average DM density in the universe, and our semi-major axis $a_2$ ensures that the inner binary lies well within the DM spike. In the spike model with fixed $\gamma_{\mathrm{sp}}$, a lower DM density means a greater distance and a longer external orbital period, which is actually beneficial for the occurrence of flips after the same number of periods due to the greatly extended dynamical friction action time.



\section{Conclusion and discussion}

This work derives for the first time the evolution equations (Appendix~\ref{app:octupole}) for a hierarchical triple system up to the octupole order without eliminating ascending nodes. Meanwhile, by considering the important roles played by Brown's Hamiltonian term and the dynamical friction of a DM spike, we find that the dynamical friction from a DM spike could indeed flip this hierarchical triple system within a large range of dark matter density.

In addition, as the spike index $\gamma_{\mathrm{sp}}$ increases (DM density increases), the number of flips would increase over a period of time; as both the inner and outer semi-major axes increase while keeping their ratio $a_1/a_2$ fixed (DM density decreases), the number of flips would also increase within the same number of outer orbital periods due to the greatly extended dynamical friction action time. In such cases, the flip can act as a promising probe for DM via not only electromagnetic waves but also gravitational radiations.

Furthermore, we also find that Brown's Hamiltonian term can bring in a much larger orbital eccentricity to the system, but this correspondence between inclination and eccentricity is slightly broken, although the flip brought by dynamical friction still corresponds to an extremely large eccentricity. Therefore, this coincides with our previous expectation that the dynamical-friction promotion of eccentricity can increase the probability of occurrences of flips.



In future studies, one can calculate and directly compare the GW signals~\cite{Naoz:2012bx} with and without a DM spike, and detect this GW signal difference in future space-based GW observatories such as LISA, Taiji, and TianQin. In addition, if both stars of the inner binary are pulsars, the electromagnetic wave signal is also worth studying to put stringent constraints on the DM density, see Ref.~\cite{Pani:2015qhr} for more details.

\acknowledgments

We sincerely thank Kaye Jiale Li for the helpful discussions. We particularly thank an anonymous referee and the handling editor for professional suggestions that greatly improved the quality of the manuscript.
This work is supported by 
the National Key Research and Development Program of China Grants No. 2021YFA0718304, No. 2021YFC2203004, and No. 2020YFC2201501,
the National Natural Science Foundation of China Grants No. 12422502, No. 12105344, No. 12235019,  No. 12047503, No. 12073088, No. 11821505, No. 11991052, and No. 11947302,
the Strategic Priority Research Program of the Chinese Academy of Sciences (CAS) Grant No. XDB23030100, No. XDA15020701, the Key Research Program of the CAS Grant No. XDPB15,  the Key Research Program of Frontier Sciences of CAS,
the Science Research Grants from the China Manned Space Project with No. CMS-CSST-2021-B01 (supported by China Manned Space Program through its Space Application System).

\appendix
\section{DM sipke parameters $\rho_\mathrm{sp}$ and $R_\mathrm{sp}$}\label{Spike}

Here we show how to derive empirical radius $R_\mathrm{sp}$ and normalization constant $\rho_\mathrm{sp}$ from the object masses $m_1$, $m_2$, $m_3$ and redshift $z$ step by step.

Firstly, one may define the central black mass $M_\mathrm{BH}$ as the third body mass $m_3$ or the total mass $m_1+m_2+m_3$. In our cases, $m_3$ is always much greater than $m_1+m_2$, so the two definitions are basically equivalent. The redshift $z$ used in the calculation is fixed to $z=1$ for simplicity in this paper.

Then, one can obtain the stellar mass $M_*$ via~\cite{Kormendy:2013dxa}
\begin{align}
M_\mathrm{BH}=4.9\times10^8\,M_\odot(M_\mathrm{bulge}/10^{11}\,M_\odot)^{1.17}\,,
\end{align}
where the bulge mass $M_\mathrm{bulge}=0.615M_*$~\cite{Chen:2018znx}, $M_\odot$ represents solar mass.

After that, the relation between the dark matter halo mass $M_{200}$ and stellar mass $M_*$ takes this form~\cite{Girelli:2020goz}
\begin{align}
\frac{M_*}{M_{200}}(z)=\frac{2A_\mathrm{sh}(z)}{\Big[\big(\frac{M_{200}}{M_\mathrm{sh,A}(z)}\big)^{-\beta_\mathrm{sh}(z)}+\big(\frac{M_{200}}{M_\mathrm{sh,A}(z)}\big)^{\gamma_\mathrm{sh}(z)}\Big]}\,,
\end{align}
where $\lg M_\mathrm{sh}(z)=B_\mathrm{sh}+z\,\cdot\mu_\mathrm{sh}$, $A_\mathrm{sh}(z)=C_\mathrm{sh}\cdot(1+z)^{\nu_\mathrm{sh}}$, $\gamma_\mathrm{sh}(z)=D_\mathrm{sh}\cdot(1+z)^{\eta_\mathrm{sh}}$, $\beta_\mathrm{sh}(z)=E_\mathrm{sh}+z\,\cdot F_\mathrm{sh}$, and we use the best fit parameters in Tab.~3 of \cite{Girelli:2020goz}: $B_\mathrm{sh}=11.79$, $C_\mathrm{sh}=0.046$, $D_\mathrm{sh}=0.709$, $E_\mathrm{sh}=0.96$, $F_\mathrm{sh}=0.043$, $\mu_\mathrm{sh}=0.20$,
$\nu_\mathrm{sh}=-0.38$, and $\eta_\mathrm{sh}=-0.18$.

Next, the critical density $\rho_{c}(z)$ can be given in terms of redshift $z$
\begin{align}
\rho_c(z)\equiv\frac{3H_0^2}{8\pi G}\cdot[(1-\Omega_\mathrm{m})+\Omega_\mathrm{m}(1+z)^3]\,,
\end{align}
where $H_0=67.27\,\mathrm{km\, s^{-1}\, Mpc^{-1}}$ and $\Omega_\mathrm{m}=0.3166$~\cite{Planck:2018vyg}.

Immediately, $R_{200}$ can be obtained by substituting the above $\rho_{c}(z)$ into the relation $M_{200}\equiv200\times\frac43\pi\rho_c(z)R_{200}^3$.

In addition, the concentration $c_{200}$ can be expressed as
\begin{align}
c_{200}=C_c(z)\Big(\frac{M_{200}}{10^{12}\frac{100}{H_0}M_\odot}\Big)^{-\gamma_c(z)}\Big[1+\big(\frac{M_{200}}{M_{c,0}(z)}\big)^{0.4}\Big]\,,
\end{align}
where these functions $C_c(z)$, $\gamma_c(z)$, $M_{c,0}(z)$ can be obtained by constructing interpolation functions from the data in Tab.~2~\cite{Klypin:2014kpa}.

Hereafter, the scale radius $r_s$ can be solved since $c_{200}$ is also the ratio of $R_{200}$ to $r_s$ ($c_{200}=R_{200}/r_s$). Then $\rho_s$ can be solved from
\begin{align}
M_{200}=\int_0^{R_{200}}4\pi r^2\rho_\mathrm{NFW}(r)\mathrm{d}r\,,
\end{align}
where $\rho_\mathrm{NFW}(r)$ is the Navarro-Frenk-White (NFW) profile~\cite{Navarro:1996gj} and can be shown as an illustrative example,
\begin{align}
\rho_\mathrm{NFW}(r)=\frac{\rho_s}{(r/r_s)(1+r/r_s)^{2}}.
\end{align}

Finally, the empirical radius $R_\mathrm{sp}$ can be obtained with
\begin{align}
R_\mathrm{sp}=a_\gamma r_s\sqrt{M_\mathrm{BH}/(\rho_sr_s^3)}\,,
\end{align}
where $a_\gamma\simeq0.1$~\cite{Gondolo:1999ef}.
Then, from the matching condition $\rho_\mathrm{sp}=\rho_\mathrm{spike}(R_\mathrm{sp})=\rho_\mathrm{NFW}(R_\mathrm{sp})$, the normalization constant $\rho_\mathrm{sp}$ can be fixed.

\section{EOM of octupole Hamiltonian}\label{app:octupole}

First, we define $U=(4-3B+24e_1^2)$, $P=-4+B-10e_1^2$, $Q=\left[4-3B+(-4+B)e_1^2\right]\cos(2i_1)$ to simplify subsequent expressions. The evolution of the variables for both inner and outer orbits are 

\begin{align}\label{EOM}
\dot{e}_1&=\alpha \beta^{e_1}\left[C^{e_1}_0+C^{e_1}_1\cos(\Omega)+C^{e_1}_2\cos(2\Omega)+C^{e_1}_3\cos(3\Omega)+C^{e_1}_4\sin(\Omega)+C^{e_1}_5\sin(2\Omega)+C^{e_1}_6\sin(3\Omega)\right]\,,
\end{align}
where

\begin{align}
\beta^{e_1}&=\frac{(m_1+m_2)^{1/2}(1-e_1^2)^{1/2}}{m_1m_2e_1(Ga_1)^{1/2}}\,,
\nonumber\\
C^{e_1}_0&=6\cos(g_1)\sin(g_2)\sin(i_1)\left[16-5B+12e_1^2+5B\cos(2i_1)\right]\left[\sin(i_2)+5\sin(3i_2)\right]\,,
\nonumber\\
C^{e_1}_1&=-2\sin(g_1)\cos(g_2)\left[36-15B+132e_1^2-5U\cos(2i_1)\right]\left[3+5\cos(2i_2)\right]
\nonumber\\
&\quad+\cos(g_1)\sin(g_2)\cos(i_1)\left[16-15B+12e_1^2+15B\cos(2i_1)\right]\left[\cos(i_2)+15\cos(3i_2)\right]\,,
\nonumber\\
C^{e_1}_2&=10\sin(i_1)\{16U\sin(g_1)\cos(g_2)\cos(i_1)\sin(2i_2)
\nonumber\\
&\quad-\cos(g_1)\sin(g_2)\left[-16+9B-12e_1^2+3B\cos(2i_1)\right]\left[\sin(i_2)-3\sin(3i_2)\right]\}\,,
\nonumber\\
C^{e_1}_3&=60\sin^2(i_2)\{\sin(g_1)\cos(g_2)\left[-12+5B-44e_1^2-U\cos(2i_1)\right]
\nonumber\\
&\quad+\cos(g_1)\sin(g_2)\cos(i_1)\cos(i_2)\left[16-7B+12e_1^2-B\cos(2i_1)\right]\}\,,
\nonumber\\
C^{e_1}_4&=-2\cos(g_1)\cos(g_2)\cos(i_1)\left[16-15B+12e_1^2+15B\cos(2i_1)\right]\left[3+5\cos(2i_2)\right]
\nonumber\\
&\quad+\sin(g_1)\sin(g_2)\left[3\times(-12+5B-44e_1^2)+5U\cos(2i_1)\right]\left[\cos(i_2)+15\cos(3i_2)\right]\,,
\nonumber\\
C^{e_1}_5&=80\sin(i_1)\sin(i_2)\{\cos(g_1)\cos(g_2)\cos(i_2)\left[16-9B+12e_1^2-3B\cos(2i_1)\right]
\nonumber\\
&\quad+U\sin(g_1)\sin(g_2)\cos(i_1)\left[1+3\cos(2i_2)\right]\}\,,
\nonumber\\
C^{e_1}_6&=-60\sin^2(i_2)\{-\cos(g_1)\cos(g_2)\cos(i_1)\left[-16+7B-12e_1^2+B\cos(2i_1)\right]
\nonumber\\
&\quad+\sin(g_1)\sin(g_2)\cos(i_2)\left[12-5B+44e_1^2+U\cos(2i_1)\right]\}\,,
\nonumber
\end{align}

\begin{equation}
\begin{aligned}
&\frac{{\rm d}i_1}{{\rm d}t}=\alpha \beta^{i_1}\left[C^{i_1}_0+C^{i_1}_1\cos(\Omega)+C^{i_1}_2\cos(2\Omega)+C^{i_1}_3\cos(3\Omega)+C^{i_1}_4\sin(\Omega)+C^{i_1}_5\sin(2\Omega)+C^{i_1}_6\sin(3\Omega)\right]\,,
\end{aligned}
\end{equation}
where

\begin{align}
\beta^{i_1}&=\frac{(m_1+m_2)^{1/2}}{m_1m_2(Ga_1)^{1/2}(1-e_1^2)^{1/2}}\,,
\nonumber\\
C^{i_1}_0&=-6\cos(g_1)\sin(g_2)\cos(i_1)\left[16-5B+12e_1^2+5B\cos(2i_1)\right]\left[\sin(i_2)+5\sin(3i_2)\right]\,,
\nonumber\\
C^{i_1}_1&=\sin(i_1)\{-40P\sin(g_1)\cos(g_2)\cos(i_1)\left[3+5\cos(2i_2)\right]
\nonumber\\
&\quad+\cos(g_1)\sin(g_2)\left[16+5B+12e_1^2+15B\cos(2i_1)\right]\left[\cos(i_2)+15\cos(3i_2)\right]\}\,,
\nonumber\\
C^{i_1}_2&=160P\sin(g_1)\cos(g_2)\cos(2i_1)\sin(2i_2)
\nonumber\\
&\quad+10\cos(g_1)\sin(g_2)\cos(i_1)\left[-16+B-12e_1^2+3B\cos(2i_1)\right]\left[\sin(i_2)-3\sin(3i_2)\right]\,,
\nonumber\\
C^{i_1}_3&=60\sin(i_1)\sin^2(i_2)\{4P\sin(g_1)\cos(g_2)\cos(i_1)
\nonumber\\
&\quad+\cos(g_1)\sin(g_2)\cos(i_2)\left[16-3B+12e_1^2-B\cos(2i_1)\right]\}\,,
\nonumber\\
C^{i_1}_4&=-2\cos(g_1)\cos(g_2)\sin(i_1)\left[16+5B+12e_1^2+15B\cos(2i_1)\right]\left[3+5\cos(2i_2)\right]
\nonumber\\
&\quad-10P\sin(g_1)\sin(g_2)\sin(2i_1)\left[\cos(i_2)+15\cos(3i_2)\right]\,,
\nonumber\\
C^{i_1}_5&=40\cos(g_1)\cos(g_2)\cos(i_1)\sin(2i_1)\left[B-4(4+3e_1^2)+3B\cos(2i_1)\right]
\nonumber\\
&\quad-40P\sin(g_1)\sin(g_2)\cos(2i_1)\left[\sin(i_2)-3\sin(3i_2)\right]\,,
\nonumber\\
C^{i_1}_6&=-60\sin(i_1)\sin^2(i_2)\{\cos(g_1)\cos(g_2)\left[16-3B+12e_1^2-B\cos(2i_1)\right]
\nonumber\\
&\quad-4P\sin(g_1)\sin(g_2)\cos(i_1)\cos(i_2)\}\,,
\nonumber
\end{align}

\begin{equation}
\begin{aligned}\label{g1dot}
&\frac{{\rm d}g_1}{{\rm d}t}=\alpha \beta^{g1}\left[C^{g1}_0+C^{g1}_1\cos(\Omega)+C^{g1}_2\cos(2\Omega)+C^{g1}_3\cos(3\Omega)+C^{g1}_4\sin(\Omega)+C^{g1}_5\sin(2\Omega)+C^{g1}_6\sin(3\Omega)\right]\,,
\end{aligned}
\end{equation}
where
\begin{align}
\beta^{g1}&=\frac{(m_1+m_2)^{1/2}}{m_1m_2e_1^2(Ga_1)^{1/2}(1-e_1^2)^{1/2}}\,,
\nonumber\\
C^{g_1}_0&=\frac{3}{2}\sin(g_1)\sin(g_2)\csc(i_1)\left[\sin(i_2)+5\sin(3i_2)\right]\{-32+15B-4(22+5B)e_1^2
\nonumber\\
&\quad+176e_1^4+4\cos(2i_1)\left[8-5B+(38+5B)e_1^2-32e_1^4\right]+5B\cos(4i_1)\}\,,
\nonumber\\
C^{g_1}_1&=2\cos(g_1)\cos(g_2)\left[5B(3-5e_1^2)+36(-1+e_1^4)+5Q\right]\left[3+5\cos(2i_2)\right]
\nonumber\\
&\quad-\sin(g_1)\sin(g_2)\cos(i_1)\left[16-15B+4(29+5B)e_1^2-104e_1^4+15B\cos(2i_1)\right]\left[\cos(i_2)+15\cos(3i_2)\right]\,,
\nonumber\\
C^{g_1}_2&=-80\cos(g_1)\cos(g_2)\cot(i_1)\sin(2i_2)\{Q-\left[(-4+3B)\times(-1+e_1^2)\right]\}
\nonumber\\
&\quad+\frac{5}{2}\sin(g_1)\sin(g_2)\csc(i_1)\left[\sin(i_2)-3\sin(3i_2)\right]\{-32+15B-4(22+5B)e_1^2
\nonumber\\
&\quad+176e_1^4+4\left[8-3B+(22+B)e_1^2-16e_1^4\right]\cos(2i_1)-3B\cos(4i_1)\}\,,
\nonumber\\
C^{g_1}_3&=60\sin^2(i_2)\{\cos(g_1)\cos(g_2)(-12+5B-3Be_1^2+12e_1^4-Q)
\nonumber\\
&\quad+\sin(g_1)\sin(g_2)\cos(i_1)\cos(i_2)\left[-16+7B-4(13+B)e_1^2+40e_1^4+B\cos(2i_1)\right]\}\,,
\nonumber\\
C^{g_1}_4&=2\sin(g_1)\cos(g_2)\cos(i_1)\left[16-15B+4(29+5B)e_1^2-104e_1^4+15B\cos(2i_1)\right]\left[3+5\cos(2i_2)\right]
\nonumber\\
&\quad+\cos(g_1)\sin(g_2)\left[5B(3-5e_1^2)+36(-1+e_1^4)+5Q\right]\left[\cos(i_2)+15\cos(3i_2)\right]\,,
\nonumber\\
C^{g_1}_5&=20\csc(i_1)\sin(i_2)\{\sin(g_1)\cos(g_2)\cos(i_2)\left[-32+15B-4(22+5B)e_1^2\right.
\nonumber\\
&\quad+176e_1^4+4(8-3B+22e_1^2+Be_1^2-16e_1^4)\cos(2i_1)-3B\cos(4i_1)\left.\right]
\nonumber\\
&\quad-2\cos(g_1)\sin(g_2)\cos(i_1)\left[Q-(4-3B-4e_1^2+3Be_1^2)\right]\left[1+3\cos(2i_2)\right]\}\,,
\nonumber\\
C^{g_1}_6&=-30\sin^2(i_2)\cos(g_2)\cos(i_1)\{\left[-4-e_1^2+12e_1^4+(4+17e_1^2)\cos(2i_1)\right]\sin(g_1)
\nonumber\\
&\quad-7e_1^2\left[7-4e_1^2+\cos(2i_1)\right]\sin(3g_1)\}-60\sin^2(i_2)\cos(g_1)\sin(g_2)\cos(i_2)
\nonumber\\
&\quad\times(12-5B+3Be_1^2-12e_1^4+Q)\,,
\nonumber
\end{align}

\begin{equation}
\begin{aligned}\label{h1dot}
&\frac{{\rm d}h_1}{{\rm d}t}=\alpha \beta^{h_1}\left[C^{h_1}_0+C^{h_1}_1\cos(\Omega)+C^{h_1}_2\cos(2\Omega)+C^{h_1}_3\cos(3\Omega)+C^{h_1}_4\sin(\Omega)+C^{h_1}_5\sin(2\Omega)+C^{h_1}_6\sin(3\Omega)\right]\,,
\end{aligned}
\end{equation}
where
\begin{align}
\beta^{h_1}&=\frac{(m_1+m_2)^{1/2}}{m_1m_2(Ga_1)^{1/2}(1-e_1^2)^{1/2}}\,,
\nonumber\\
C^{h_1}_0&=6\sin(g_1)\sin(g_2)\cot(i_1)(4+5B-32e_1^2-5A)\left[\sin(i_2)+5\sin(3i_2)\right]\,,
\nonumber\\
C^{h_1}_1&=40B\cos(g_1)\cos(g_2)\cos(i_1)\left[3+5\cos(2i_2)\right]
\nonumber\\
&\quad+\sin(g_1)\sin(g_2)(36+5B-8e_1^2+15A)\left[\cos(i_2)+15\cos(3i_2)\right]\,,
\nonumber\\
C^{h_1}_2&=\frac{5}{4}\csc(i_1)\{-128B\cos(g_1)\cos(g_2)\cos(2i_1)\sin(2i_2)
\nonumber\\
&\quad+8\sin(g_1)\sin(g_2)\cos(i_1)(-12+B-16e_1^2+3A)\left[\sin(i_2)-3\sin(3i_2)\right]\}\,,
\nonumber\\
C^{h_1}_3&=60\sin^2(i_2)\left[-4B\cos(g_1)\cos(g_2)\cos(i_1)+\sin(g_1)\sin(g_2)\cos(i_2)(4-3B+24e_1^2-A)\right]\,,
\nonumber\\
C^{h_1}_4&=-2\sin(g_1)\cos(g_2)(36+5B-8e_1^2+15A)\left[3+5\cos(2i_2)\right]
\nonumber\\
&\quad+20B\cos(g_1)\sin(g_2)\cos(i_1)\left[\cos(i_2)+15\cos(3i_2)\right]\,,
\nonumber\\
C^{h_1}_5&=80\csc(i_1)\sin(i_2)\{\sin(g_1)\cos(g_2)\cos(i_1)\cos(i_2)(-12+B-16e_1^2+3A)
\nonumber\\
&\quad-B\cos(g_1)\sin(g_2)\cos(2i_1)\left[1+3\cos(2i_2)\right]\}\,,
\nonumber\\
C^{h_1}_6&=60\sin^2(i_2)\left[\sin(g_1)\cos(g_2)(-4+3B-24e_1^2+A)-4B\cos(g_1)\sin(g_2)\cos(i_1)\cos(i_2)\right]\,,
\nonumber
\end{align}

\begin{equation}
\begin{aligned}\label{e2dot}
&\frac{{\rm d}e_2}{{\rm d}t}=\alpha \beta^{e_2}\left[C^{e_2}_0+C^{e_2}_1\cos(\Omega)+C^{e_2}_2\cos(2\Omega)+C^{e_2}_3\cos(3\Omega)+C^{e_2}_4\sin(\Omega)+C^{e_2}_5\sin(2\Omega)+C^{e_2}_6\sin(3\Omega)\right]\,,
\end{aligned}
\end{equation}
where
\begin{align}
\beta^{e_2}&=\frac{(m_1+m_2+m_2)^{1/2}(1-e_2^2)^{1/2}}{(m_1+m_2)m_3e_2(Ga_2)^{1/2}}\,,
\nonumber\\
C^{e_2}_0&=-2\sin(g_1)\cos(g_2)\sin(i_1)(-28+5B-56e_1^2-5A)\left[\sin(i_2)+5\sin(3i_2)\right]\,,
\nonumber\\
C^{e_2}_1&=\sin(g_1)\cos(g_2)\cos(i_1)(-4-5B+32e_1^2+5A)\left[\cos(i_2)+15\cos(3i_2)\right]
\nonumber\\
&\quad-2\cos(g_1)\sin(g_2)\left[16-5B+12e_1^2+5B\cos(2i_1)\right]\left[3+5\cos(2i_2)\right]\,,
\nonumber\\
C^{e_2}_2&=5\sin(i_1)\{-32B\cos(g_1)\sin(g_2)\cos(i_1)\sin(2i_2)
\nonumber\\
&\quad-2\sin(g_1)\cos(g_2)(-4+3B-24e_1^2+A)\left[\sin(i_2)-3\sin(3i_2)\right]\}\,,
\nonumber\\
C^{e_2}_3&=20\sin^2(i_2)\{\sin(g_1)\cos(g_2)\cos(i_1)\cos(i_2)(20-7B+64e_1^2-A)
\nonumber\\
&\quad+\cos(g_1)\sin(g_2)\left[-16+5B-12e_1^2+3B\cos(2i_1)\right]\}\,,
\nonumber\\
C^{e_2}_4&=\cos(g_1)\cos(g_2)\left[16-5B+12e_1^2+5B\cos(2i_1)\right]\left[\cos(i_2)+15\cos(3i_2)\right]
\nonumber\\
&\quad+2\sin(g_1)\sin(g_2)\cos(i_1)(-4-5B+32e_1^2+5A)\left[3+5\cos(2i_2)\right]\,,
\nonumber\\
C^{e_2}_5&=80\sin(i_1)\sin(i_2)\{B\cos(g_1)\cos(g_2)\cos(i_1)\left[1+3\cos(2i_2)\right]
\nonumber\\
&\quad+\sin(g_1)\sin(g_2)\cos(i_2)(-4+3B-24e_1^2+A)\}\,,
\nonumber\\
C^{e_2}_6&=5\sin^2(i_2)\{4\cos(g_1)\cos(g_2)\cos(i_2)\left[16-5B+12e_1^2-3B\cos(2i_2)\right]
\nonumber\\
&\quad-4\sin(g_1)\sin(g_2)\cos(i_1)(-20+7B-64e_1^2+A)\}\,,
\nonumber
\end{align}

\begin{equation}
\begin{aligned}\label{i2dot}
&\frac{{\rm d}i_2}{{\rm d}t}=\alpha \beta^{i_2}\left[C^{i_2}_0+C^{i_2}_1\cos(\Omega)+C^{i_2}_2\cos(2\Omega)+C^{i_2}_3\cos(3\Omega)+C^{i_2}_4\sin(\Omega)+C^{i_2}_5\sin(2\Omega)+C^{i_2}_6\sin(3\Omega)\right]\,,
\end{aligned}
\end{equation}
where
\begin{align}
\beta^{i_2}&=\frac{(m_1+m_2+m_2)^{1/2}}{(m_1+m_2)m_3e_2(Ga_2)^{1/2}(1-e_2^2)^{1/2}}\,,
\nonumber\\
C^{i_2}_0&=2\sin(g_1)\cos(g_2)\sin(i_1)(-28+5B-56e_1^2-5A)\left[11\cos(i_2)+5\cos(3i_2)\right]\,,
\nonumber\\
C^{i_2}_1&=2\sin(i_2)\{\sin(g_1)\cos(g_2)\cos(i_1)(-4-5B+32e_1^2+5A)\left[13+15\cos(2i_2)\right]
\nonumber\\
&\quad+20\cos(g_1)\sin(g_2)\cos(i_2)(16-5B+12e_1^2+5B\cos(2i_1))\}\,,
\nonumber\\
C^{i_2}_2&=-20\sin(i_1)\{\sin(g_1)\cos(g_2)\cos(i_2)(-4+3B-24e_1^2+A)\left[-7+3\cos(2i_2)\right]
\nonumber\\
&\quad+16B\cos(g_1)\sin(g_2)\cos(i_1)\cos(2i_2)\}\,,
\nonumber\\
C^{i_2}_3&=5\sin(i_2)\{2\sin(g_1)\cos(g_2)\cos(i_1)(-20+7B-64e_1^2+A)\left[-5+\cos(2i_2)\right]
\nonumber\\
&\quad+8\cos(g_1)\sin(g_2)\cos(i_2)\left[-16+5B-12e_1^2+3B\cos(2i_1)\right]\}\,,
\nonumber\\
C^{i_2}_4&=2\sin(i_2)\{\cos(g_1)\cos(g_2)\left[16-5B+12e_1^2+5B\cos(2i_1)\right]\left[13+15\cos(2i_2)\right]
\nonumber\\
&\quad-20\sin(g_1)\sin(g_2)\cos(i_1)\cos(i_2)(-4-5B+32e_1^2+5A)\}\,,
\nonumber\\
C^{i_2}_5&=10\sin(i_1)\{8B\cos(g_1)\cos(g_2)\cos(i_1)\cos(i_2)\left[7-3\cos(2i_2)\right]
\nonumber\\
&\quad+8\sin(g_1)\sin(g_2)\cos(2i_2)\left[-4+3B-24e_1^2+A\right]\}\,,
\nonumber\\
C^{i_2}_6&=10\sin(i_2)\{\cos(g_1)\cos(g_2)\left[-16+5B-12e_1^2+3B\cos(2i_1)\right]\left[-5+\cos(2i_2)\right]
\nonumber\\
&\quad-4\sin(g_1)\sin(g_2)\cos(i_1)\cos(i_2)(-20+7B-64e_1^2+A)\}\,,
\nonumber
\end{align}

\begin{equation}
\begin{aligned}\label{g2dot}
&\frac{{\rm d}g_2}{{\rm d}t}=\alpha \beta^{g_2}\left[C^{g_2}_0+C^{g_2}_1\cos(\Omega)+C^{g_2}_2\cos(2\Omega)+C^{g_2}_3\cos(3\Omega)+C^{g_2}_4\sin(\Omega)+C^{g_2}_5\sin(2\Omega)+C^{g_2}_6\sin(3\Omega)\right]\,,
\end{aligned}
\end{equation}
where
\begin{align}
\beta^{g_2}&=\frac{(m_1+m_2+m_2)^{1/2}}{(m_1+m_2)m_3e_2^2(Ga_2)^{1/2}(1-e_2^2)^{1/2}}\,,
\nonumber\\
C^{g_2}_0&=\{\sin(g_1)\sin(g_2)\sin(i_1)\csc(i_2)(-28+5B-56e_1^2-5A)
\nonumber\\
&\quad\times\left[1+3e_2^2+4\cos(2i_2)-5(1+7e_2^2)\cos(4i_2)\right]\}\,,
\nonumber\\
C^{g_2}_1&=-2\cos(g_1)\cos(g_2)\left[16-5B+12e_1^2+5B\cos(2i_1)\right]\left[3+22e_2^2+5(1+6e_2^2)\cos(2i_2)\right]
\nonumber\\
&\quad-2\sin(g_1)\sin(g_2)\cos(i_1)\cos(i_2)(-4-5B+32e_1^2+5A)\left[-7-5e_2^2+15(1+7e_2^2)\cos(2i_2)\right]\,,
\nonumber\\
C^{g_2}_2&=5\sin(i_1)\{32B\cos(g_1)\cos(g_2)\cos(i_1)\cot(i_2)\left[-1-4e_2^2+(1+6e_2^2)\cos(2i_2)\right]
\nonumber\\
&\quad+\sin(g_1)\sin(g_2)\csc(i_2)(-4+3B-24e_1^2+A)
\nonumber\\
&\quad\times\left[1+3e_2^2-4(1+2e_2^2)\cos(2i_2)+3(1+7e_2^2)\cos(4i_2)\right]\}\,,
\nonumber\\
C^{g_2}_3&=-10\{\cos(g_1)\cos(g_2)\left[-16+5B-12e_1^2+3B\cos(2i_1)\right]\left[-1-2e_2^2+(1+6e_2^2)\cos(2i_2)\right]
\nonumber\\
&\quad\sin(g_1)\sin(g_2)\cos(i_1)\cos(i_2)(-20+7B-64e_1^2+A)\left[-1-3e_2^2+(1+7e_2^2)\cos(2i_2)\right]\}\,,
\nonumber\\
C^{g_2}_4&=2\sin(g_1)\cos(g_2)\cos(i_1)(-4-5B+32e_1^2+5A)\left[3+22e_2^2+5(1+6e_2^2)\cos(2i_2)\right]
\nonumber\\
&\quad-2\cos(g_1)\sin(g_2)\cos(i_2)\left[16-5B+12e_1^2+5B\cos(2i_1)\right]\left[-7-5e_2^2+15(1+7e_2^2)\cos(2i_2)\right]\,,
\nonumber\\
C^{g_2}_5&=-20\sin(i_1)\{2\sin(g_1)\cos(g_2)\cot(i_2)(-4+3B-24e_1^2+A)\left[-1-4e_2^2+(1+6e_2^2)\cos(2i_2)\right]
\nonumber\\
&\quad-B\cos(g_1)\sin(g_2)\cos(i_1)\csc(i_2)\left[1+3e_2^2-4(1+2e_2^2)\cos(2i_2)+3(1+7e_2^2)\cos(4i_2)\right]\}\,,
\nonumber\\
C^{g_2}_6&=-10\{\sin(g_1)\cos(g_2)\cos(i_1)(20-7B+64e_1^2-A)\left[-1-2e_2^2+(1+6e_2^2)\cos(2i_2)\right]
\nonumber\\
&\quad+\cos(g_1)\sin(g_2)\cos(i_2)\left[5B-4(4+3e_1^2)+3B\cos(2i_1)\right]\left[-1-3e_2^2+(1+7e_2^2)\cos(2i_2)\right]\}\,,
\nonumber
\end{align}

\begin{equation}
\begin{aligned}\label{h2dot}
&\frac{{\rm d}h_2}{{\rm d}t}=\alpha \beta^{h_2}\left[C^{h_2}_0+C^{h_2}_1\cos(\Omega)+C^{h_2}_2\cos(2\Omega)+C^{h_2}_3\cos(3\Omega)+C^{h_2}_4\sin(\Omega)+C^{h_2}_5\sin(2\Omega)+C^{h_2}_6\sin(3\Omega)\right]\,,
\end{aligned}
\end{equation}
where
\begin{align}
\beta^{h_2}&=\frac{(m_1+m_2+m_2)^{1/2}}{(m_1+m_2)m_3(Ga_2)^{1/2}(1-e_2^2)^{1/2}}\,,
\nonumber\\
C^{h_2}_0&=2\sin(g_1)\sin(g_2)\sin(i_1)\csc(i_2)(-28+5B-56e_1^2-5A)\left[\cos(i_2)+15\cos(3i_2)\right]\,,
\nonumber\\
C^{h_2}_1&=\csc(i_2)\{20\cos(g_1)\cos(g_2)\sin(2i_2)\left[16-5B+12e_1^2+5B\cos(2i_1)\right]
\nonumber\\
&\quad+\sin(g_1)\sin(g_2)\cos(i_1)(-4-5B+32e_1^2+5A)\left[\sin(i_2)+45\sin(3i_2)\right]\}\,,
\nonumber\\
C^{h_2}_2&=5\sin(i_1)\csc(i_2)\{-64B\cos(g_1)\cos(g_2)\cos(i_1)\cos(2i_2)
\nonumber\\
&\quad+2\sin(g_1)\sin(g_2)(-4+3B-24e_1^2+A)\left[\cos(i_2)-9\cos(3i_2)\right]\}\,,
\nonumber\\
C^{h_2}_3&=40\cos(g_1)\cos(g_2)\cos(i_2)\left[-16+5B-12e_1^2+3B\cos(2i_1)\right]
\nonumber\\
&\quad+10\sin(g_1)\sin(g_2)\cos(i_1)(-20+7B-64e_1^2+A)\left[1+3\cos(2i_2)\right]\,,
\nonumber\\
C^{h_2}_4&=40\sin(g_1)\cos(g_2)\cos(i_1)\cos(i_2)(4+5B-32e_1^2-5A)
\nonumber\\
&\quad+2\cos(g_1)\sin(g_2)\left[16-5B+12e_1^2+5B\cos(2i_1)\right]\left[23+45\cos(2i_2)\right]\,,
\nonumber\\
C^{h_2}_5&=40\sin(i_1)\csc(i_2)\{2\sin(g_1)\cos(g_2)\cos(2i_2)(-4+3B-24e_1^2+A)
\nonumber\\
&\quad+B\cos(g_1)\sin(g_2)\cos(i_1)\left[\cos(i_2)-9\cos(3i_2)\right]\}\,,
\nonumber\\
C^{h_2}_6&=40\sin(g_1)\cos(g_2)\cos(i_1)\cos(i_2)(20-7B+64e_1^2-A)
\nonumber\\
&\quad+10\cos(g_1)\sin(g_2)\left[-16+5B-12e_1^2+3B\cos(2i_1)\right]\left[1+3\cos(2i_2)\right]\nonumber\,,
\end{align}




\bibliographystyle{JHEP}
\bibliography{ref.bib}

\end{document}